\PassOptionsToPackage{sort&compress}{natbib}
\documentclass[final,5p,times,twocolumn]{elsarticle}

\usepackage{amssymb}
\usepackage{amsmath}
\usepackage{xcolor}
\usepackage{lipsum}
\usepackage{cuted}
\usepackage{bm}

\journal{Physics Open}

\begin{document}

\begin{frontmatter}



\title{Thermalization-induced self-polarization of temporally incoherent light:\\ 
A nonequilibrium wave turbulence kinetic approach}


\author[first]{L. Zanaglia}
\affiliation[first]{organization={Department of Physics, University of Calabria},
            addressline={Via Pietro Bucci}, 
            city={Rende},
            postcode={87036}, 
            country={Italy}}

\author[six]{J. Garnier}
\affiliation[six]{organization={CMAP, CNRS, Ecole Polytechnique, Institut Polytechnique de Paris},
            city={Palaiseau},
            postcode={91120}, 
            country={France}}

\author[first]{M. Ferraro}

\author[second]{T. Torres}
\affiliation[second]{organization={Universit\'e Bourgogne Europe, CNRS, Laboratoire Interdisciplinaire Carnot de Bourgogne ICB UMR 6303},
            city={Dijon},
            postcode={21000}, 
            country={France}}

\author[third]{C. Michel}
\affiliation[third]{organization={Universit\'e C\^ote d’Azur, CNRS, Institut de Physique de Nice},
            city={Nice},
            country={France}}

\author[third]{V. Doya}

\author[second]{G. Millot}

\author[four]{S. Wabnitz}
\affiliation[four]{organization={Department of Information Engineering, Electronics, and Telecommunications, Sapienza University of Rome},
            addressline={Via Eudossiana 18}, 
            city={Rome},
            postcode={00184}, 
            country={Italy}}

\author[five]{I. Carusotto}
\affiliation[five]{organization={INO-CNR Pitaevskii BEC Center and Dipartimento di Fisica, Universit\'a di Trento},
            city={Povo},
            postcode={38123}, 
            country={Italy}}

\author[second]{A. Picozzi}

\begin{abstract}
Conservative nonlinear waves generally undergo an irreversible process of thermalization toward the Rayleigh--Jeans (RJ) equilibrium state. Light thermalization has recently been experimentally investigated in the two-dimensional transverse spatial dynamics of multimode waveguides. Here, we consider the one-dimensional temporal dynamics of incoherent light propagating in a weakly birefringent single-mode fiber. We show that light thermalization is responsible for a self-polarization of the incoherent wave during its propagation. Within the framework of wave turbulence theory, we show that this phenomenon originates from a nonequilibrium process of adiabatic cooling: polarization fluctuations are continuously transferred toward the high-frequency tails of a local RJ equilibrium state, whose effective temperature decreases during propagation, thereby increasing the degree of light polarization. Moreover, in the absence of a high-frequency cut-off that would regularize the ultraviolet catastrophe in the temporal domain, the optical wave exhibits an ever-increasing degree of polarization, ultimately approaching unity. These results show that it is thermodynamically advantageous for the optical field to evolve toward a polarized state, which provides new insight into spontaneous self-organization phenomena in conservative Hamiltonian systems operating far from thermodynamic equilibrium.
\end{abstract}




\end{frontmatter}




\section{Introduction}
\label{introduction}

Non-integrable Hamiltonian systems of random waves generally undergo thermalization -- an irreversible evolution toward thermodynamic equilibrium, characterized by a state of maximum entropy. In the weakly nonlinear regime, this process is accurately described by the wave turbulence (WT) theory \cite{zakharov92,Newell01,zakharov04,nazarenko11,Newell_Rumpf,galtier22,onorato23}, which has been successfully applied to various physical systems \cite{lvov97,rumpf09,nazarenko25,nazarenko17,onorato20,onorato15}, including optical waves propagating in liquid crystals \cite{laurie12}, or in optical fibers \cite{babin07,turitsyn13,churkin15,pitois06,barviau09,PRL10,PR14,lagrange07}. In spite of the formal reversibility of the Hamiltonian system, the wave turbulence kinetic equation describes the actual irreversible evolution to the Rayleigh-Jeans (RJ) equilibrium distribution, as expressed by a $H$-theorem of entropy growth.

Optical systems provide a natural setting for exploring the emergence of collective phenomena from nonlinear wave interactions. In this context, there is increasing interest in the study of so-called `fluids of light' with effective photon-photon interactions, which constitute the photonic analogue of atomic Bose gases \cite{carusotto13,bloch22,glorieux23,glorieux25,wrightNP22,ferraroAPX23}. The notion of a fluid of light arises from a mathematical correspondence between the propagation of light through a nonlinear medium and the conservative time evolution of a quantum fluid of interacting photons~\cite{larre15}. Propagating fluids of light have consequently provided a versatile platform for investigating a broad range of fascinating phenomena, including the generation of superfluid Bogoliubov sound waves \cite{fontaine18,michel18}, complex vortex dynamics \cite{fleischerBKT,michel21,glorieux23b,congy24,pavloff25}, binary fluids~\cite{Piekarski25}, spin-orbit-coupled fluids~\cite{martone21,martone23}, analogue gravity and cosmology \cite{segev15,braidotti22,marcucci19,roger16,marino19,glorieux22,shah23}, turbulent cascade regimes \cite{ferreira22,rasooli23}, and the dynamical emergence of prethermalized equilibrium states \cite{berges04,larre16,santic18,abuzarli22}.

Among the phenomena associated with the propagation of fluids of light, complete three-dimensional quantum thermalization to a Bose-Einstein distribution and photon condensation in bulk Kerr media were predicted as early as 2016 \cite{chiocchetta16}. Nevertheless, because collisional scattering processes between single photons are exceedingly rare in standard media, these phenomena were expected to require impractically long propagation distances. In contrast, for classical fields, thermalization and condensation are predicted to occur much more rapidly owing to bosonic stimulation effects, as discussed in~\cite{davis01,PRL05,blakie08,krstulovic11,nazarenko11,PR14}.

In this context, the thermalization of classical optical waves toward the RJ equilibrium distribution has become a topic of significant interest in recent years. In particular, by considering quasi-monochromatic optical beams -- thus neglecting temporal dynamics and restricting the analysis to the transverse spatial degrees of freedom -- the problem reduces to a two-dimensional (2D) spatial thermalization process. Within this framework,  thermalization toward the classical 2D RJ distribution has been predicted for optical waves propagating in multimode fibers \cite{aschieri11} using the framework of wave turbulence theory \cite{zakharov92,Newell01,nazarenko11,Newell_Rumpf,PR14}. The finite number of guided modes in a multimode fiber not only accelerates the convergence to the RJ equilibrium, but also regularizes the ultraviolet catastrophe that is intrinsic to classical-wave systems  \cite{aschieri11}. It is precisely in this guided-wave geometry that RJ thermalization has recently been investigated experimentally through the transverse 2D spatial dynamics of a quasi-monochromatic speckle beam propagating in a multimode fiber \cite{PRL19,PRL20,EPL21,pourbeyram22,mangini22,podivilov22,mangini23}. This phenomenon is closely connected to the spatial beam self-cleaning effect \cite{wright16,krupa17}, whose underlying physical mechanism has been analyzed in several theoretical studies \cite{PRL19,podivilov19,PRA19,christodoulides19,sidelnikov19,parto19,makris20,makris22,baudinOC23}; see also the review articles \cite{ferraro23,ferraro25_rev}. 
More broadly, this sustained research activity has contributed to the emergence of the field of optical thermodynamics \cite{christodoulides19,wright22,kottos20,kottos23,ren23,efremidis24_press,efremidis24,ren25,efremidis26,kabat26}. Within this rapidly developing area, several important advances have been achieved, including the theoretical prediction \cite{christodoulides19} and experimental observation of negative-temperature equilibrium states \cite{PRL23,muniz23,ferraro25}, the calorimetry of photon gases \cite{ferraro24}, the observation of their Joule–Thomson expansion \cite{kirsch25}, and the development of non-equilibrium thermodynamic formalisms \cite{PRL22,lu24,Pyrialakos24,kottos24,kottos24_defect,liu25,efremidis25}.

From a different perspective, temporally incoherent light waves propagating in single-mode fibers have been considered for a long time to study optical turbulence  \cite{babin07,turitsyn13,churkin15,pitois06,barviau09,PRL10,PR14}. In particular, a recent work has reported experimental evidence of an incipient process of optical thermalization in the temporal domain by considering light propagation in a weakly birefringent single-mode fiber \cite{torres26}. In this setting, the nonequilibrium process of thermalization was shown to be characterized by the self-polarization of the incoherent optical wave, i.e., by an increase of its degree of polarization ${\cal P}$ during propagation through the fiber. Our aim in this article is to provide further theoretical insight into this process of self-polarization  induced by natural thermalization. For this purpose, we follow the approach recently developed to describe spatio-temporal thermalization~\cite{zanaglia26}. The analysis shows that light self-polarization is driven by a nonequilibrium process of adiabatic cooling of the optical field. Specifically, in the absence of a global RJ equilibrium in the temporal domain, the wave turbulence kinetic equations, combined with numerical simulations, reveal that the optical field adiabatically follows a local RJ equilibrium, characterized by a temperature that progressively decreases along the propagation. This effective cooling results from a continuous transfer of polarization fluctuations toward the high-frequency tails of the optical spectrum, thereby increasing the degree of polarization. Furthermore, since no high-frequency cut-off is available to regularize the ultraviolet catastrophe, the optical field exhibits an ever-increasing degree of polarization, ultimately approaching unity.

Beyond its intrinsic fundamental interest, this self-polarization process may provide a route toward a universal polarizer capable of repolarizing unpolarized incoherent light with high efficiency, overcoming the intrinsic 50\% power loss associated with conventional polarization-selection schemes \cite{mandel}. As a matter of fact, universal polarizers for coherent light have previously been proposed through counterpropagating nonlinear interactions in single-mode fibers \cite{fatome12,pitois05,guasoni14,guasoni16}, and were subsequently extended to the multimode-fiber regime \cite{ji23}. The mechanism of self-polarization reported here is fundamentally different: it originates from the thermalization of the optical field, a process that naturally involves incoherent light, and therefore enables the self-polarization of incoherent light.

\section{NLS model}
\label{sec:nls}

We consider a system of two coherently coupled NLS equations, which is known as a generic model for the description of vector phenomena in various physical contexts.
In the context of binary Bose-Einstein condensates (BEC), the model describes a two-component spinor BEC  with an external field driving the population transfer between the two atomic levels. 
Coherently coupled BECs are important, e.g., to study collective excitations with Faraday patterns \cite{cominotti22}, paramagnetic to ferromagnetic quantum phase transitions \cite{cominotti23},
or metastable bubble nucleation in ferromagnetic superfluids \cite{zenesini23}. 

In optics, the coherently coupled NLS equations are known to describe the polarization dynamics of an optical wave propagating in a weakly birefringent optical fiber \cite{millot_prl97,millot_josab98,agrawal}: 
\begin{eqnarray}
&&\hspace*{-0.2in} 
i\partial_z u_x = - \beta \partial_{tt} u_x + \alpha u_x + \gamma (|u_x|^2+ \kappa |u_y|^2)u_x + \gamma \rho u_x^* u_y^2
\label{eq:nlsu_x}\\
&&\hspace*{-0.2in} 
i\partial_z u_y  = - \beta \partial_{tt} u_y - \alpha u_y + \gamma (|u_y|^2+ \kappa |u_x|^2)u_y  + \gamma \rho u_y^* u_x^2
\label{eq:nlsu_y}
\end{eqnarray}
where $u_{x,y}$ denote the rectilinear polarization components of the optical field. As usual in optics, the distance $z$ of propagation plays the role of an evolution `time' variable, while $t$ denotes the retarded time in a reference frame moving with the waves. 
The parameter $\gamma$ denotes the nonlinear Kerr coefficient, $\beta$ is the group-velocity dispersion, while $\kappa$ and $\rho$ are dimensionless interaction coefficients -- we will consider the case $\kappa=2\rho$ relevant to our optical system. The 
parameter $\alpha$ originates from the weak birefringence of the optical fiber \cite{agrawal}, and it can be assumed positive without loss of generality, $\alpha >0$.
For simplicity, we restrict our analysis to waves propagating in the normal-dispersion regime $\beta>0$, while the corresponding analysis for $\beta<0$ follows along similar lines.
The dispersion relations of the waves read
\begin{eqnarray}
k_{x}(\omega) = \beta \omega^2 + \alpha, \quad k_{y}(\omega) = \beta \omega^2 - \alpha.
\label{eq:k_xy}
\end{eqnarray}
where $\omega$ is the frequency (see~\ref{app:kinetic} for details).

In the following, we consider an incoherent optical wave that exhibits fluctuations that are statistically stationary in time. 
Denoting by $\langle \cdot \rangle$ an average over the realizations of the random initial conditions $u_{x,y}(z=0,t)$, the power defined by 
\begin{eqnarray}
N =  \langle |u_x|^2 \rangle + \langle |u_y|^2 \rangle .
\label{eq:N}
\end{eqnarray}
is a conserved quantity: it is independent of $t$ by statistical stationarity, and independent of $z$ by the NLS Eq.(\ref{eq:nlsu_x}-\ref{eq:nlsu_y}).
Note that, if we consider a time window ${\cal T}$ that is much larger than the correlation time of light $\tau_c$, then time-averaged quantities are equivalent to statistical averages, so that we have
\begin{eqnarray}
\frac{1}{\cal T}\int_0^{\cal T} |u_x|^2 +|u_y|^2 dt \simeq N,
\end{eqnarray}
when ${\cal T} \gg \tau_c$.

The NLS equation also conserves the Hamiltonian $H = E+U$, whose linear component 
\begin{eqnarray}
E=E_K + E_\alpha,
\label{eq:E_cE_k}
\end{eqnarray}
has a kinetic contribution 
\begin{eqnarray}
E_{K}= \langle |\partial_t u_x|^2 \rangle  + \langle |\partial_t u_y|^2 \rangle,
\label{eq:E_K}
\end{eqnarray}
and a birefringence contribution
\begin{eqnarray}
E_{\alpha}= \alpha \big(\langle |u_x|^2\rangle-\langle|u_y|^2\rangle \big),
\label{eq:E_C}
\end{eqnarray}
while the nonlinear component reads: 
\begin{eqnarray}
U = \gamma \Big( \frac{1}{2}  \langle|u_x|^4\rangle + \frac{1}{2} \langle|u_y|^4\rangle + \kappa  \langle|u_x|^2 |u_y|^2 \rangle \Big) \nonumber \\  
+  \frac{\gamma \rho}{2}  \Big(  \langle u_x^{*2} u_y^{2}\rangle+  \langle u_x^{2} u_y^{*2}\rangle \Big).
\label{eq:U}
\end{eqnarray}

An important quantity that characterizes the incoherent optical wave is the degree of polarization \cite{mandel}:
\begin{eqnarray}
{\cal P}(z)=\Big(1-4 {\rm det}({\bf J})/{\rm tr}({\bf J})^2 \Big)^{1/2},
\label{eq:P}
\end{eqnarray}
where ${\bf J}_{ij}=\langle u_i u_j^* \rangle$ is the 2$\times$2 coherence matrix, with `det' and `tr' the determinant and trace operations. The degree of polarization ${\cal P}$ is bounded by 0 and 1, which correspond to unpolarized, and fully polarized, waves, respectively.

We note that, in addition to the effect of self-polarization induced by thermalization, the recent work reported in Ref.\cite{torres26} has identified a distinct regime in which strong phase correlations between the incoherent waves $u_x$ and $u_y$ spontaneously emerge, giving rise to a reversible oscillatory dynamics of the incoherent waves. In this article, we do not consider this oscillatory regime mediated by anomalous correlators, see Refs.\cite{lvov94,guasoni17,zaleski20,villois25}. Specifically, throughout this work, we assume that the threshold condition for the emergence of anomalous correlations is not satisfied, namely that $\alpha > \gamma (1-\kappa/2) (N_y-N_x)$, where $N_{x,y}=\langle |u_{x,y}|^2 \rangle$ \cite{torres26,note}. 
Thus, in the regime considered here, self-polarization reflects a redistribution of population between the polarization components, rather than the emergence of coherent phase correlations between them.


\section{Wave turbulence kinetic equations}
\label{sec:KE}

The process of self-polarization takes place far from thermal equilibrium, so that its understanding requires a nonequilibrium approach. On the basis of the wave turbulence theory, we derive in the \ref{app:kinetic} the  kinetic equations governing the evolutions of the averaged spectra $\langle {\tilde u}_{\nu}(\omega,z){\tilde u}_{\nu}^*(\omega',z)\rangle = 2\pi n_\nu(\omega,z) \delta(\omega-\omega')$ of the waves, where ${\tilde u}_\nu(\omega,z)$ denotes the Fourier transform of ${u}_\nu(t,z)$, for $\nu=x,y$. 

The kinetic equations describe the spectral evolution of the waves even arbitrarily far from thermal equilibrium, provided that the dynamics remains weakly nonlinear throughout the propagation, $|U| \ll |E|$. 
The collision terms provide a kinetic description of the nonlinear terms in the vector NLS Eq.(\ref{eq:nlsu_x}-\ref{eq:nlsu_y}):
\begin{eqnarray}
\partial_z n_{x}(\omega,z) = \kappa^2 {\cal C}oll_I[n_{x},n_{y}] + \rho^2 {\cal C}oll_C[n_{x},n_{y}], 
\label{eq:collx} \\
\partial_z n_{y}(\omega,z) = \kappa^2 {\cal C}oll_I[n_{y},n_{x}] + \rho^2 {\cal C}oll_C[n_{y},n_{x}].
\label{eq:colly} 
\end{eqnarray}
The collision terms are rather cumbersome, they are reported in explicit form in \ref{app:kinetic}. Let us briefly comment on their physical kinetic interpretation. The term ${\cal C}oll_I[{\bm n}]$ corresponds to the second nonlinear term in the NLS Eqs.(\ref{eq:nlsu_x}-\ref{eq:nlsu_y}). It describes the incoherent cross-interaction as a collisional gas of `particles' polarized along $x$ and $y$, $(x,y) \to (x,y)$. Accordingly, this collision term conserves the number of particles $N_{x}$ and $N_{y}$ in each polarization component. The term ${\cal C}oll_C[{\bm n}]$ originates from the third nonlinear term in the NLS Eqs. (\ref{eq:nlsu_x}-\ref{eq:nlsu_y}). It describes a phase-sensitive polarization coupling and has a distinct physical interpretation: each elementary interaction corresponds to the annihilation of two $x$-polarized particles and the creation of two $y$-polarized particles, $(x,x)\to(y,y)$, or vice versa. This collision term is responsible for an exchange of `particles' among $u_x$ and $u_y$, so that the kinetic Eqs.(\ref{eq:collx}-\ref{eq:colly}) only conserves the total power 
\begin{eqnarray}
N = \sum_\nu \frac{1}{2\pi}\int n_\nu(\omega,z) d\omega,
\label{eq:power}
\end{eqnarray}
where the sum is carried over $\nu=x,y$. Finally, note that the collision term associated with the first nonlinear term in the NLS Eqs.(\ref{eq:nlsu_x}-\ref{eq:nlsu_y}) vanishes identically. This property results from the degeneracy of the resonances inherent to the strictly one-dimensional system considered here, see \citep{OE11,zakharov04}.

In the one dimensional geometry considered here, the kinetic equations can be simplified. In the \ref{app:kinetic} we derive the following form of the kinetic equations for the vector NLS Eq.(1-2):
\begin{strip}
\begin{eqnarray}
&&\hspace*{-0.2in} 
\partial_z n_{x}(\omega,z) = \frac{\kappa^2 \gamma^2}{4\pi \beta} \int  
 n_{x}({\omega'}) \, n_{y}(\omega)\,n_{y}({\omega'})\,n_{x}(\omega)  \left( n_{x}^{-1}(\omega)+n_{y}^{-1}({\omega'})-n_{y}^{-1}(\omega)-n_{x}^{-1}({\omega'})\right)  \frac{d{\omega'}}{|{\omega'}-\omega|} \nonumber\\
&&\hspace*{0.4in} 
+ \frac{ \rho^2 \gamma^2}{2\pi \beta} \sum_s \int_0^{\infty}   
 n_{y}\left(\omega + s\sqrt{\zeta} {\eta}\right) 
 \, n_{y}\left(\omega-s\sqrt{\zeta} /{\eta} \right)\,n_{x}\left(\omega + s\sqrt{\zeta}( {\eta}-1/{\eta}) \right) \, n_{x}(\omega) \nonumber\\ 
&&\hspace*{1.6in}
\times \left[ n_{x}^{-1}(\omega)+n_{x}^{-1}\left(\omega +s\sqrt{\zeta} ({\eta}- 1/{\eta})\right)-n_{y}^{-1} \left(\omega +s\sqrt{\zeta} {\eta}\right)-n_{y}^{-1}\left(\omega-s\sqrt{\zeta}  / {\eta}\right)\right] \frac{d{\eta}}{{\eta}} , 
\label{eq:kin_n_x} \\
&&\hspace*{-0.2in} 
\partial_z n_{y}(\omega,z) = \frac{\kappa^2 \gamma^2}{4\pi \beta} \int  
 n_{y}({\omega'}) \, n_{x}(\omega)\,n_{x}({\omega'})\,n_{y}(\omega)  
 \left( n_{x}^{-1}(\omega)+n_{y}^{-1}({\omega'}) -n_{y}^{-1}(\omega)-n_{x}^{-1}({\omega'})\right)  \frac{d{\omega'}}{|{\omega'}-\omega|} \nonumber\\
&&\hspace*{0.4in} 
+ \frac{ \rho^2 \gamma^2}{2\pi \beta} \sum_s \int_0^{\infty}   
 n_{x}\left(\omega + s\sqrt{\zeta} {\eta}\right) 
 \, n_{x}\left(\omega+s\sqrt{\zeta} /{\eta} \right)
 \, n_{y}\left(\omega +s \sqrt{\zeta} ({\eta}+1 /{\eta}) \right)
 \, n_{y}(\omega) \nonumber\\ 
&&\hspace*{1.6in}
\times \left[ n_{y}^{-1}(\omega)+n_{y}^{-1}\left(\omega +s\sqrt{\zeta}( {\eta}+1  / {\eta})\right)-n_{x}^{-1} \left(\omega +s\sqrt{\zeta} {\eta}\right)-n_{x}^{-1}\left(\omega+s\sqrt{\zeta}  / {\eta}\right)\right] \frac{d{\eta}}{{\eta}} , 
\label{eq:kin_n_y}
\end{eqnarray}
where $\zeta=2\alpha/\beta>0$ and $\sum_{s}$ denotes a sum over $s=+1$ and $s=-1$. Note that, in Eqs.(\ref{eq:kin_n_x}-\ref{eq:kin_n_y}), the first (second) term  refers to the incoherent (coherent) collision term ${\cal C}oll_I[{\bm n}]$ (${\cal C}oll_C[{\bm n}]$) introduced in Eqs.(\ref{eq:collx}-\ref{eq:colly}), respectively.
As commented above, the kinetic equations conserve the total power $N=N_x+N_y$, with $N_\nu=\frac{1}{2\pi} \int n_{\nu}(\omega,z) d\omega$ and $\nu=x,y$. More explicitly, 
we obtain from the kinetic Eq.(\ref{eq:kin_n_x}-\ref{eq:kin_n_y}) the following expression for the evolution of the powers along the polarization components (see \ref{app:kinetic_power}):
\begin{eqnarray}
&&\hspace*{-0.2in}
d_z N_{x}(z) = \frac{2\rho^2 \gamma^2}{(2\pi)^2 \beta}  \int d\omega  \int d\theta 
 n_{y}(\omega+\sqrt{\zeta}  \cosh \theta) \, n_{y}(\omega -\sqrt{\zeta}  \cosh \theta)\,n_{x}(\omega +\sqrt{\zeta}  \sinh \theta)\,n_{x}(\omega-\sqrt{\zeta}  \sinh \theta) \nonumber\\ 
&&\hspace*{1.2in}
\times \left[ n_{x}^{-1} \big(\omega +\sqrt{\zeta}  \sinh \theta \big)+n_{x}^{-1} \big(\omega -\sqrt{\zeta}  \sinh \theta\big)-n_{y}^{-1}\big(\omega +\sqrt{\zeta}  \cosh \theta\big)-n_{y}^{-1} \big(\omega - \sqrt{\zeta}  \cosh \theta \big)\right] ,
\label{eq:dzNx} \\ 
&&\hspace*{-0.2in}
d_z N_{y}(z) =  - \partial_z N_{x}(z)   .
\label{eq:dzNy}
\end{eqnarray}
\end{strip}
The kinetic equations also conserve the linear energy 
\begin{eqnarray}
E = \sum_\nu \frac{1}{2\pi}\int k_\nu(\omega) n_\nu(\omega,z) d\omega,
\label{eq:energy}
\end{eqnarray}
with $k_\nu(\omega)$ given by (\ref{eq:k_xy}) and $\nu=x,y$, which dominates the nonlinear energy in the considered weakly nonlinear regime, $|E/U| \gg 1$. 

At variance with the NLS Eq.(\ref{eq:nlsu_x}-\ref{eq:nlsu_y}) that is formally reversible, the kinetic equation is irreversible, as expressed by a $H$-theorem of entropy growth, $d{\cal S}/dz \geq 0$, where 
\begin{eqnarray}
{\cal S}(z) = \sum_\nu \frac{1}{2\pi}\int \log\big({n}_\nu(\omega,z)\big) d\omega
\label{eq:ent}
\end{eqnarray}
is the nonequilibrium entropy.


\section{Self-polarization effect}
\label{sec:repolarization}

In this section, we report numerical evidence of the self-polarization of temporally incoherent light. In the weakly nonlinear regime, we report simulations of both NLS Eq.(\ref{eq:nlsu_x}-\ref{eq:nlsu_y}) and kinetic Eq.(\ref{eq:kin_n_x}-\ref{eq:kin_n_y}), showing self-polarization and nonequilibrium thermalization. We then consider the strongly nonlinear regime, where WT theory is no longer applicable, and therefore consider exclusively numerical simulations of the NLS equations.  

\subsection{Weakly nonlinear regime}
\label{subsec:repolar_weak}

We consider the weakly nonlinear regime underlying the wave turbulence kinetic theory presented in Sec.~\ref{sec:KE}. Specifically, we assume a highly incoherent wave field with a broad spectrum, in such a way that linear dispersion effects dominate nonlinear interactions. In this regime  $|U/E| \ll 1$, the linear energy is essentially conserved $E \simeq$~const, as described by the kinetic equations. As initial condition, we consider two uncorrelated random waves $u_{x,y}(t,z=0)$ with zero mean, Gaussian statistics and Gaussian-shaped initial spectrum 
\begin{eqnarray}
n_{\nu}^0(\omega)= \frac{\sqrt{2\pi} N_{\nu}^0}{\sigma} \exp\Big(-\frac{\omega^2}{2\sigma^2}\Big).
\label{eq:init_gauss}
\end{eqnarray}
The spectral width $\sigma$ can be chosen arbitrarily. For the following analysis, it is useful to introduce a key characteristic time, usually called `healing time' $\tau_0=\sqrt{\beta L_{nl}}$, which characterizes the temporal scale at which linear and nonlinear effects are of the same order, where $L_{nl} = 1/(\gamma N)$, denotes the nonlinear length. In the weakly nonlinear regime we have $\sigma \tau_0 \gg 1$, so that the correlation time $t_c=1/\sigma$ is much smaller than the healing time, $t_c \ll  \tau_0$.

To study the early stage of the self-polarization process, 
we substitute Eq.(\ref{eq:init_gauss}) into the equation (\ref{eq:dzNx}) that governs the evolution of $N_x(z)$.
For small propagation lengths, we derive in \ref{app:kinetic_power} the following expression for the power in the $x-$polarization component:
\begin{eqnarray}
&&\hspace*{-0.2in} 
d_z N_{x}(z) = \frac{4\rho^2 \gamma^2}{\sqrt{3} \beta \sigma^2} N_y^0 N_x^0 \exp\Big(-\frac{\zeta}{3\sigma^2}\Big) \nonumber \\
&&\hspace*{0.5in} 
\times \, K_0\Big( \frac{2\zeta}{3\sigma^2}\Big) \, \Big[ N_y^0-N_x^0 \exp\Big(\frac{2\zeta}{3\sigma^2}\Big)\Big] ,
\label{eq:init_stage}
\end{eqnarray}    
where $K_0(x)$ denotes the modified Bessel function of the second kind. Recalling that $\zeta >0$, Eq.(\ref{eq:init_stage}) shows that an equipartition of power among the polarization components $(N_x^0 = N_y^0)$ is unstable, and that a transfer of power from the $x$ to the $y$ polarization component occurs whenever 
\begin{eqnarray}
N_x^0 > N_y^0 \exp\Big(-\frac{2\zeta}{3\sigma^2}\Big).
\label{eq:stab_init}
\end{eqnarray}    
In the numerical simulations we consider the initial condition (\ref{eq:init_gauss}) with $N_x^0=N_y^0$. Accordingly, the condition (\ref{eq:stab_init}) is satisfied at $z=0$, and during the early stages of propagation, until the spectra deviate appreciably from their initial Gaussian profiles. The unidirectional transfer of power described by (\ref{eq:init_stage}) then triggers the initial stage of the self-polarization process. 

\begin{figure}[t]
	\centering 
    \includegraphics[width=0.5\textwidth]{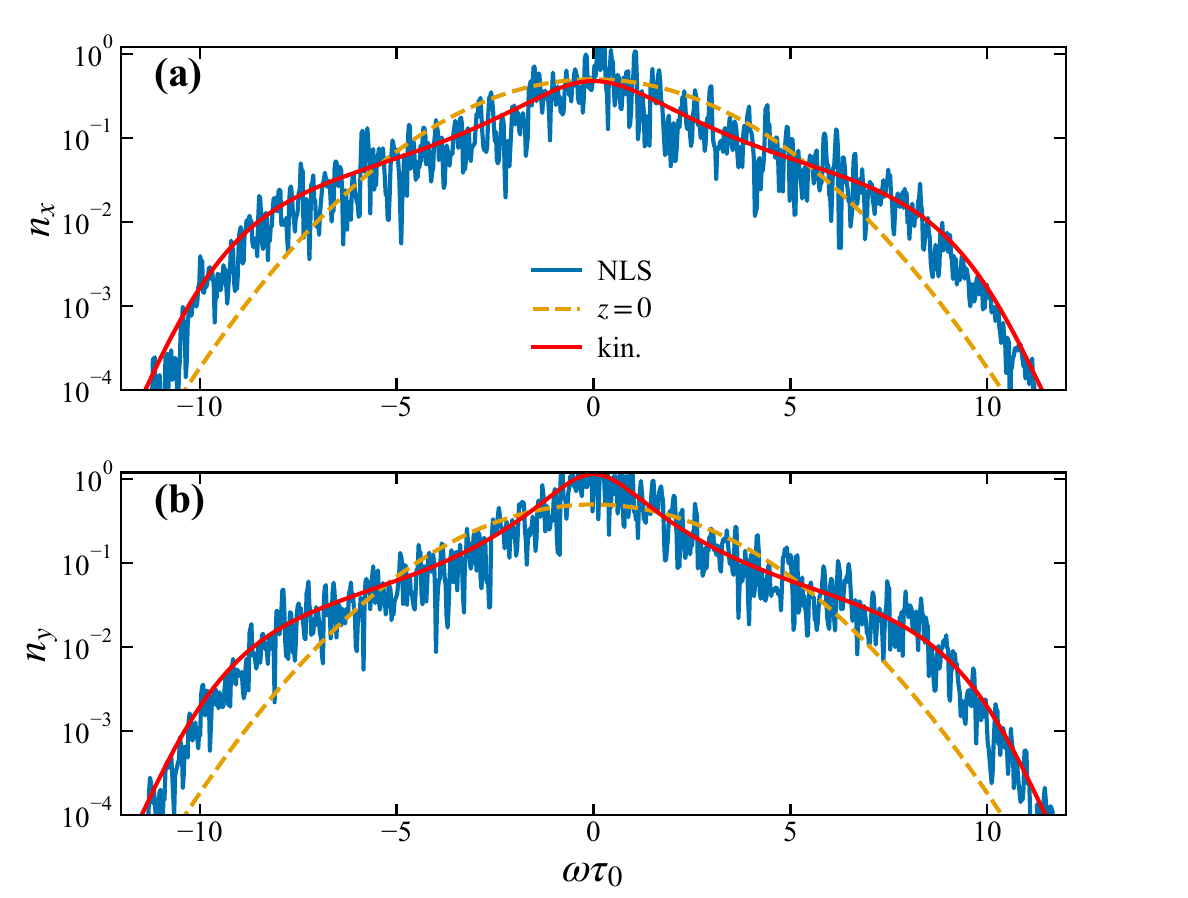}	
	\caption{\textbf{Spectra of the polarization components in the weakly nonlinear regime.} 
    Spectra of the field $n_x(\omega, z)$ (a), $n_y(\omega, z)$ (b), in semi-logarithmic scale at the propagation length $z = 17000 L_{nl}$. The orange dashed lines correspond to the Gaussian initial spectrum, see Eq.(\ref{eq:init_gauss}) with $N_x^0=N_y^0$. The blue lines report the simulation of the NLS Eqs.(\ref{eq:nlsu_x}-\ref{eq:nlsu_y}). The red lines report the simulation of the wave turbulence kinetic Eqs.(\ref{eq:kin_n_x}-\ref{eq:kin_n_y}). 
    Parameters : initial spectral width $\sigma = 0.8\pi/\tau_0$, $\alpha L_{nl} = 1, \kappa=2/3$. 
    }
	\label{fig:weak_spectra}
\end{figure}

\begin{figure}[t]
	\centering 
	\includegraphics[width=0.5\textwidth]{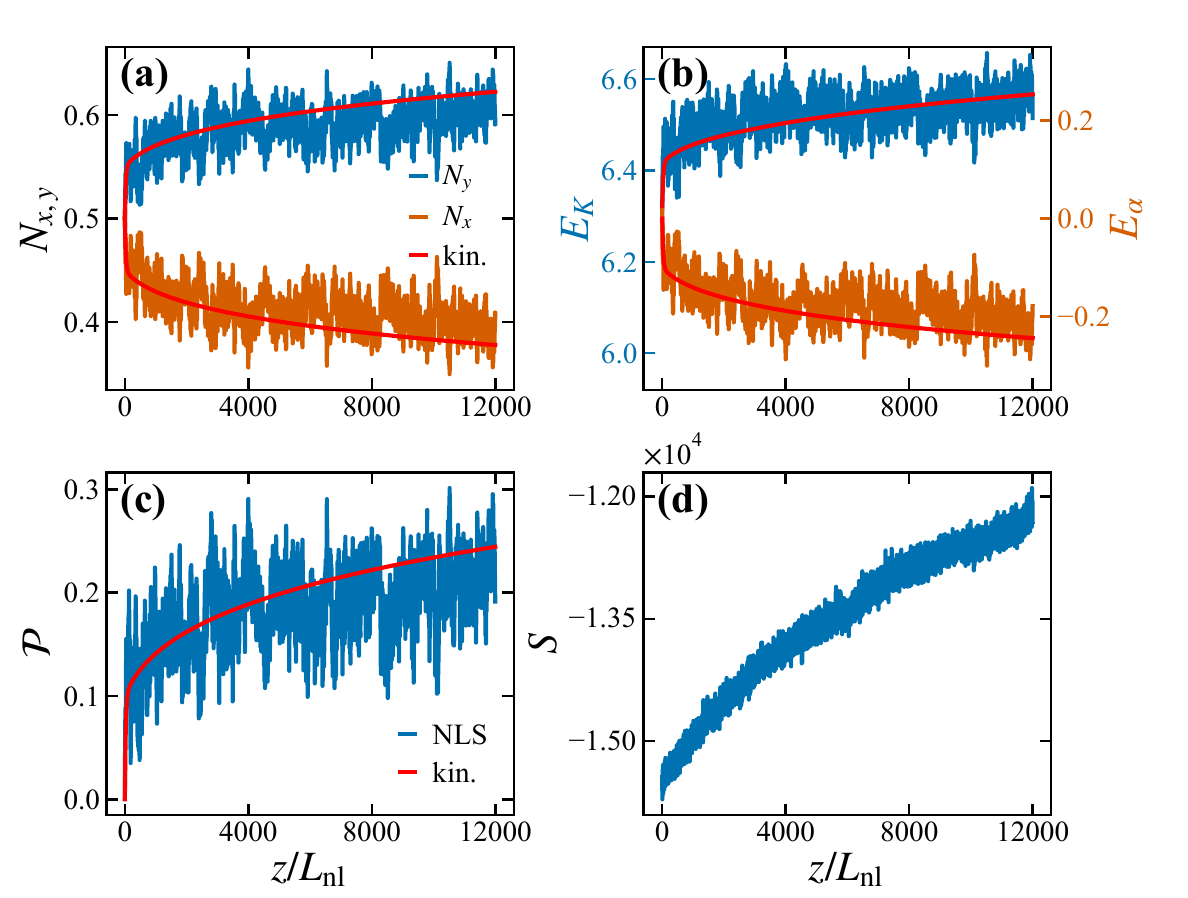}		
	\caption{\textbf{Self-polarization in the weakly nonlinear regime.} Evolution along the propagation distance $z$ of (a) the optical powers $N_y$ (dark blue curve) and $N_x$ (orange curve), (b) the kinetic energy $E_K$ (dark blue curve) and the energy $E_\alpha$ (orange curve), (c) the degree of polarization ${\cal P}$ and (d) the entropy $S$. The blue and orange lines are the result of NLS simulation of Eqs.(\ref{eq:nlsu_x}-\ref{eq:nlsu_y}). The red curves report the result of simulations of kinetic Eqs.(\ref{eq:kin_n_x}-\ref{eq:kin_n_y}). The irreversible transfer of power from the $x$ to the $y$ polarization components [panel (a)] results in an increase of the degree of polarization ${\cal P}$ [panel (c)], signaling self-polarization during propagation. Parameters: see Fig.\ref{fig:weak_spectra}. 
    } 
	\label{fig:weak_ENPS}
\end{figure}

The subsequent evolution of the system is characterized by a spectral broadening of the waves, as illustrated in Fig.~\ref{fig:weak_spectra} that reports the spectra obtained from the simulations of the NLS and kinetic equations -- notably, the good agreement between the two is achieved without any adjustable parameters. We note that the noisy NLS spectra have been smoothed using a moving average in frequency $\omega$, thereby reducing short-scale fluctuations and facilitating comparison with the kinetic simulations. Since the kinetic energy is related to the spectral width of the waves, spectral broadening results in an increase in $E_K(z)$ as the waves propagate, see Fig.~\ref{fig:weak_ENPS}(b). Unlike conventional nonlinear systems where the kinetic energy $E_K$ is the sole contribution to $E$, our system includes an additional contribution arising from the birefringence, $E=E_K+E_\alpha$. Consequently, the growth in $E_K$ is no longer constrained, as it can be offset by a corresponding decrease in $E_\alpha$. This is illustrated in Fig.~\ref{fig:weak_ENPS}(b), where $E=E_K+E_\alpha \simeq$const in the weakly nonlinear regime ($|U/E| \ll 1$). Recalling from Eq.(\ref{eq:E_C}) that $E_\alpha(z) = \alpha (N_x-N_y)$, thermalization manifests itself by an irreversible transfer of power from the $x$-axis to the $y$-axis, as illustrated in Fig.~\ref{fig:weak_ENPS}(a). This redistribution of power, in turn, gives rise to a nonvanishing degree of polarization, as shown in Fig.~\ref{fig:weak_ENPS}(c).

The analysis then indicates that the self-polarization effect is thermodynamically favorable for the optical system \cite{OE08}, as evidenced by the process of entropy growth reported in Fig.~\ref{fig:weak_ENPS}(d). It should be stressed, however, that the behavior of the entropy in Fig.~\ref{fig:weak_ENPS}(d) significantly differs from that observed when a system approaches an equilibrium state, where the entropy growth saturates to a constant value, ${\cal S}(z) \simeq$const. In contrast, here, the entropy continuously increases without sign of saturation. This aspect will be discussed in detail through the quasi-equilibrium process of adiabatic cooling in Sec.~\ref{sec:adiab_cool}.

\subsection{Strong nonlinear regime}
\label{subsec:repolar_strong}

Before analyzing further the self-polarization process in the framework of wave turbulence kinetic theory, we demonstrate in this section that self-polarization also occurs in the strongly nonlinear regime. In this regime, the linear $E$ and nonlinear $U$ contributions to the energy become comparable, so that $E$ is no longer conserved. Consequently, the assumptions underlying the wave turbulence theory are no longer satisfied, and the corresponding kinetic description cannot be applied.

\begin{figure}[t]
	\centering 
	\includegraphics[width=0.5\textwidth]{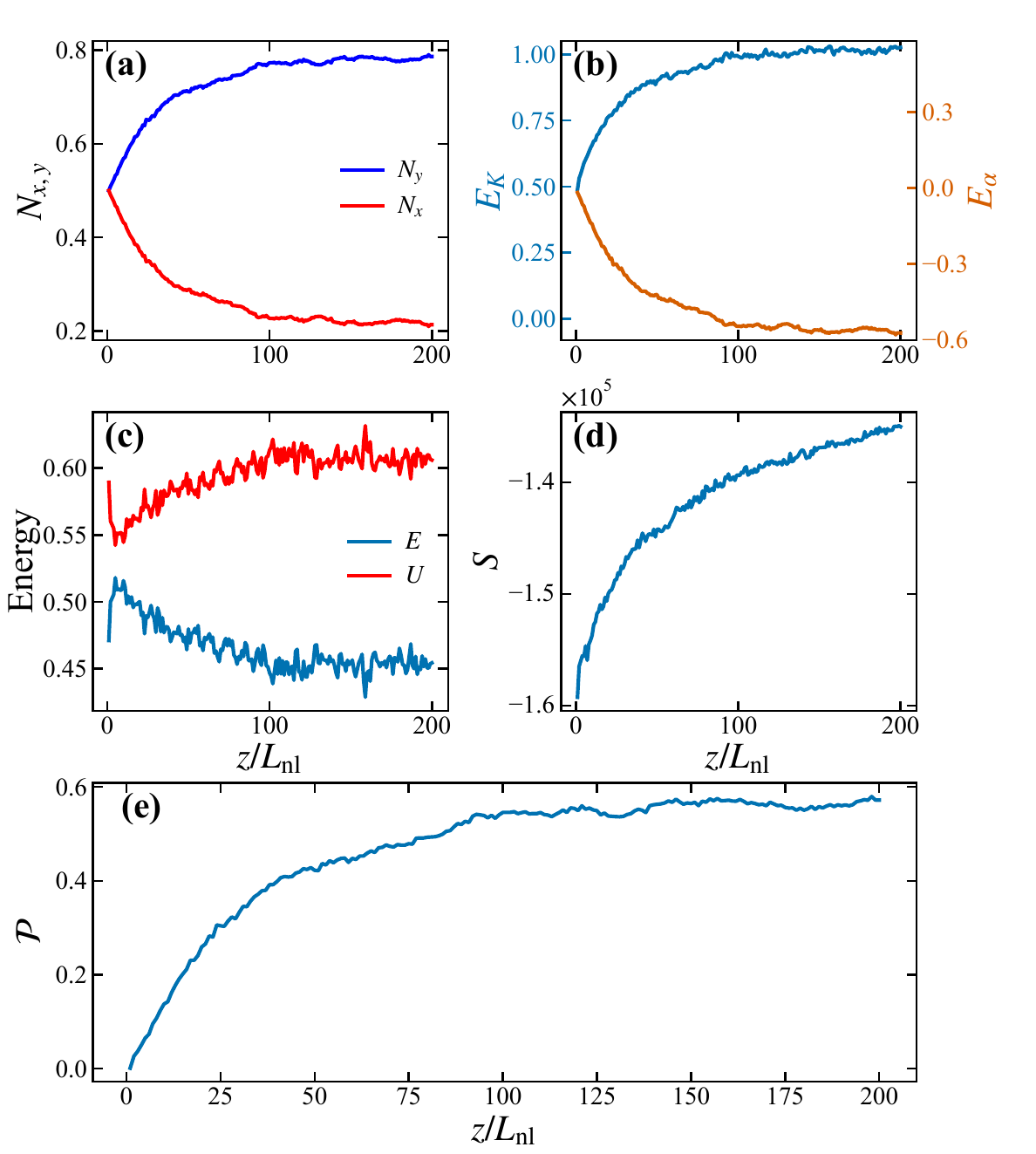}	
	\caption{\textbf{Self-polarization in the strongly nonlinear regime.} Evolution with propagation distance $z$ of (a) the optical powers $N_y$ (blue) and $N_x$ (red), (b) the kinetic energy $E_K$ (dark blue curve) and the energy $E_\alpha$ (orange curve), (c) the linear energy $E$ (blue) and nonlinear energy $U$ (red), (d) the entropy $S$, and (e) the degree of polarization $\cal{P}$. The redistribution of power among  the $x$ and $y$ polarizations in (a) is accompanied by an increase of $E_K$ and a decrease of $E_\alpha$, see panel (b). In the strongly nonlinear regime $U>E$, the linear energy $E$ is no longer conserved, as shown in (c). The entropy $S$ increases monotonically in (d), suggesting an ongoing thermalization process, as the degree of polarization $\cal{P}$ increases, see (e). Parameters are the same as in the weakly nonlinear regime in Figs.~\ref{fig:weak_spectra}-\ref{fig:weak_ENPS}, except for the initial spectral width is $\sigma=0.2\pi/\tau_0$.
} 
	\label{fig:strong_ENPS}
\end{figure}

To investigate this regime, we performed numerical simulations of the NLS Eq.(\ref{eq:nlsu_x}-\ref{eq:nlsu_y}) using the same initial state as in the weakly nonlinear case, but with narrower initial spectra. This  reduces the linear energy contribution $E$ relative to the nonlinear contribution $U$, thereby allowing us to probe the strongly nonlinear regime. Figure~\ref{fig:strong_ENPS} shows the evolution of the relevant quantities as a function of the propagation distance $z$. In contrast to the weakly nonlinear regime, where $E\simeq \mathrm{const}$, the strong nonlinear regime is characterized by substantial energy exchange between $E$ and $U$. We note in Fig.~\ref{fig:strong_ENPS}(a) that the two polarization components undergo a rapid transfer of power, with the $y$-polarized component progressively gaining power at the expense of the $x$-polarized component. This power redistribution is also reflected in the decomposition of the linear energy shown in Fig.~\ref{fig:strong_ENPS}(b): the kinetic energy $E_K$ increases with propagation, while the  energy $E_\alpha$ decreases. Although $E_K$ and $E_\alpha$ are no longer constrained to compensate each other as in the weakly nonlinear regime ($E=E_K+E_\alpha \simeq \mathrm{const}$), their evolution remain closely correlated with the transfer of power between the two polarization components. The entropy shown in panel (d) also increases monotonically throughout the propagation, with no indication of saturation, consistent with the far-from-equilibrium evolution of the system. Most of the evolution takes place over the interval $z/L_{nl}\lesssim 100$, beyond which all quantities vary much more slowly.

Figure~\ref{fig:strong_ENPS}(e) also reports the evolution of the degree of polarization ${\cal P}(z)$. As in the weakly nonlinear regime, the system undergoes a rapid initial self-polarization process. Remarkably, however, the increase in the degree of polarization is substantially larger in the strongly nonlinear regime, and the growth also seems to saturate. Note that such a pronounced growth of ${\cal P}(z)$ does not result from the development of phase correlations between the orthogonal polarization components, i.e., from non-negligible off-diagonal elements of the coherency matrix ${\bf J}$. Our numerical simulations show that even in the strongly nonlinear regime, the phase correlations remain negligible. The enhancement of the degree of polarization therefore originates almost entirely from the redistribution of power between the two polarization components, $N_x$ and $N_y$. This result suggests that the physical mechanism identified within the wave turbulence framework remains the dominant mechanism driving self-polarization in the strongly nonlinear regime.

\section{Local RJ equilibrium and adiabatic cooling}
\label{sec:adiab_cool}

\subsection{Global RJ equilibrium and ultra-violet catastrophe}

In the weakly nonlinear regime, the thermodynamic RJ equilibrium state ${n}_{\nu}^{RJ}(\omega)$ is usually obtained by maximizing the entropy ${{\cal S}}[{\bm n}]$ in (\ref{eq:ent}) under the constraints of conservation of the energy $E$ in Eq.(\ref{eq:energy}) and power $N$ in Eq.(\ref{eq:power}), which readily gives
\begin{eqnarray}
n_\nu^{RJ}(\omega)=\frac{T}{k_\nu(\omega)-\mu},
\label{eq:n_RJ}
\end{eqnarray}
where the optical temperature $T$, and optical chemical potential ${\mu}$, are the Lagrangian multipliers associated with the conservation of $E,N$, with $k_\nu(\omega)$ given by Eq.(\ref{eq:k_xy}). Note in particular that the RJ distribution (\ref{eq:n_RJ}) is a stationary solution of the kinetic equation. 

At this point, we recall that we are dealing with a closed Hamiltonian system (i.e., microcanonical statistical ensemble without a thermostat), so that $T$ and $\mu$ are not externally imposed parameters but are fully determined by the conserved quantities $(E, N)$. However, the key observation is that the integral involved in the computation of the energy in Eq.(\ref{eq:energy}) does not converge. This divergence refers to the black-body ultraviolet catastrophe, which is a well-known limitation of classical field theories in a thermal equilibrium state \cite{zakharov92,nazarenko11,blakie08}.

\subsection{Local RJ equilibrium}
\label{subsec:local_rj_eq}

The ultraviolet issue was recently addressed in Ref.~\cite{zanaglia26}, where a quasi-equilibrium process involving spatiotemporal thermalization and adiabatic cooling was identified. In the following we discuss this mechanism in the framework of the self-polarization process considered in this work. In particular, we show that spectral broadening of the waves can be understood in terms of a local RJ equilibrium, defined over a finite spectral bandwidth, with an upper cut-off frequency $\omega_c^{loc}$. This cut-off regularizes the ultraviolet divergence of the energy and provides a natural description of the locally equilibrated spectrum. More specifically, we show that, after an initial transient, the fields remain close to a local RJ equilibrium, with the cut-off frequency $\omega_c^{loc}(z)$ growing as a power law of the propagation distance $z$. At the same time, the associated local temperature $T^{loc}$ decreases according to a reciprocal power law, unveiling an intrinsic adiabatic cooling mechanism. This coupled evolution of $\omega_c^{loc}$ and $T^{loc}$ plays a central role in the polarization dynamics, and ultimately governs the long-distance behavior of the self-polarization process.

\begin{figure}[t]
	\centering 
	\includegraphics[width=0.5\textwidth]{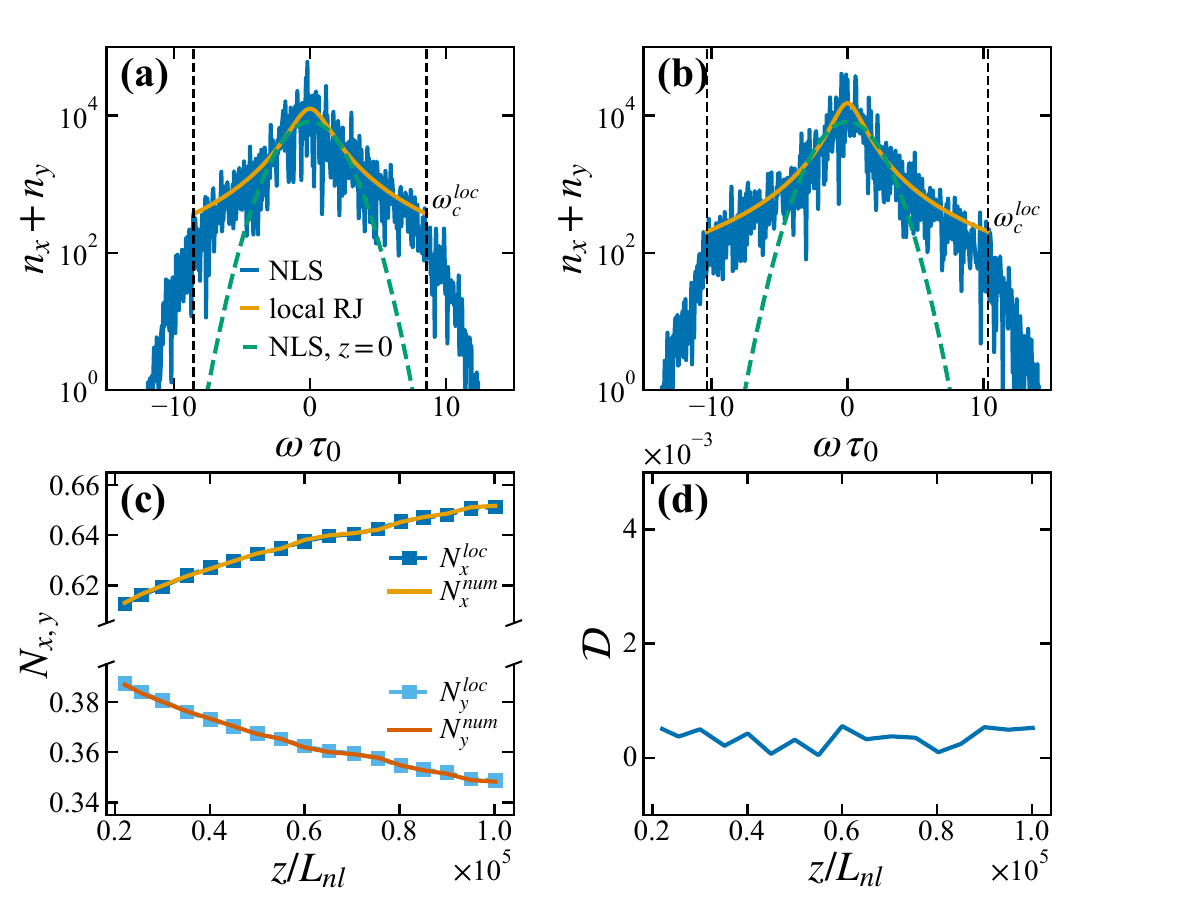}	
	\caption{\textbf{Evolution following a local equilibrium state.} Total spectrum of the field $\sum_{\nu}n_\nu(\omega,z)$ for $\nu=x,y$ at $z = 20\,000 L_{nl}$ (a), and $z = 90\,000 L_{nl}$ (b). The dashed-green lines in (a)-(b) denote the initial Gaussian spectrum. (c) Evolution during propagation of the powers associated with the local equilibrium state $N_{x,y}^{loc}(z)$ [from Eq.(\ref{eq:N_nu_loc})], together with  the powers $N_{x,y}^{num}(z)$ obtained from NLS simulations. (d) Corresponding evolution of the distance to the local RJ distribution ${\cal D}(z)$ [from Eq.(\ref{eq:dist})]. The black dashed vertical lines in (a,b) mark the positions  of the local frequency cut-off $\omega_c^{loc}$. A good agreement is observed between the numerical spectra (blue curves) and the local RJ equilibrium distribution (orange curves), over the truncated spectral window $|\omega| \le \omega_c^{loc}$, as shown in (a)-(b). This is further confirmed by the remarkable agreement between the the powers $N_{x,y}^{loc}(z)$ predicted from the local equilibrium state, and the powers $N_{x,y}^{num}(z)$ obtained in NLS simulations [panel (c)]. Consistently, the distance ${\cal D}(z)$ remains close to zero during propagation [panel (d)], indicating that the field remains close to the local RJ equilibrium distribution.
    Parameters are the same as in Fig.\ref{fig:weak_spectra}. 
    } 
	\label{fig:loc_eq1}
\end{figure}

In the numerical simulations, we start from a short-tailed Gaussian spectrum, as defined in Eq.(\ref{eq:init_gauss}) with $N_x^0=N_y^0$. After an initial transient regime, the simulations reveal that the optical field evolves close to a local equilibrium state during propagation, as illustrated in Fig.~\ref{fig:loc_eq1}(a)-(b) that reports the optical spectra at two different propagation distances. Here, the spectra of the polarization components $n_\nu(\omega)$ take the form of a truncated RJ spectrum, over a truncated frequency window $[-\omega_c^{loc}(z), \omega_c^{loc}(z)]$:
\begin{eqnarray}
n_\nu^{loc}(\omega,z)=\frac{T^{loc}(z)}{k_\nu(\omega)-\mu^{loc}(z)} \, {\bf 1}_{[-\omega_c^{loc}(z), +\omega_c^{loc}(z)]}(\omega)
\label{eq:n_loc}
\end{eqnarray}
with $T^{loc}$ and $\mu^{loc}$ the local thermodynamic parameters and ${\bf 1}_{[-\omega_c^{loc}(z), +\omega_c^{loc}(z)]}(\omega)=1$ if $\omega \in [-\omega_c^{loc}(z), +\omega_c^{loc}(z)]$ and 0 otherwise. Note that, the numerical spectra for $|\omega| \ge \omega_c^{loc}(z)$ exhibit a rapid decay, which is reminiscent of the initial Gaussian spectrum. 

It is now apparent that the quantities that characterize the local RJ equilibrium (\ref{eq:n_loc}), namely  $(T^{loc},\mu^{loc},\omega_c^{loc})$, are functions of the propagation distance $z$. Note however that, to avoid cumbersome notation, their dependence on $z$ will be left implicit in some cases. By substitution of $n_\nu^{loc}(\omega)$ given by (\ref{eq:n_loc}) into the definitions of the power we get 
\begin{eqnarray}
N_{\nu}^{loc}= \frac{T^{loc}}{\pi\sqrt{\beta(s\alpha-\mu^{loc})}}{\rm atan}\Big(\omega_c^{loc}\sqrt{\beta/(s\alpha-\mu^{loc})}\Big),
\label{eq:N_nu_loc} \\
N^{loc} = \sum_\nu N_\nu^{loc},
\label{eq:N_loc}
\end{eqnarray}
with $s=+1$ ($s=-1$) for $\nu=x$ ($\nu=y$). In a similar way, we obtain for the energy 
\begin{eqnarray}
&&\hspace*{-0.4in} 
\frac{E^{loc}}{N^{loc}} =\frac{2\omega_c^{loc}\sqrt{\beta}}{ \sum_s {\rm atan}\big(\omega_c^{loc} \sqrt{\beta/(s\alpha-\mu^{loc})}\big)/\sqrt{s\alpha-\mu^{loc}} } +\mu^{loc},
\label{eq:EsN_loc} \\
&&\hspace*{-0.4in} 
E^{loc} =2 T^{loc} \omega_c^{loc}/\pi +\mu^{loc} N^{loc},
\label{eq:E_loc}
\end{eqnarray}
where the sum is carried over $s=\pm 1$.

For a specific value of $\omega_c^{loc}$, the conserved power and energy determine a unique pair $(T^{loc},\mu^{loc})$ from Eqs.(\ref{eq:N_loc}-\ref{eq:E_loc}) -- specifically, $\mu^{loc}$ is determined by (\ref{eq:EsN_loc}), and $T^{loc}$ is subsequently fixed by (\ref{eq:E_loc}).
To compare the NLS simulations with the local RJ equilibrium (\ref{eq:n_loc}), at each propagation length $z$, we determine $(\omega_c^{loc}(z),T^{loc}(z),\mu^{loc}(z))$ by minimizing the distance to the local RJ equilibrium:
\begin{eqnarray}
{\cal D}(z)=\frac{ \sum_\nu |N_\nu^{num}(z)-N_\nu^{loc}(z)| }{ \sum_\nu N_\nu^{num}(z)+N_\nu^{loc}(z) }, \quad \nu=x,y,
\label{eq:dist}
\end{eqnarray}
where $N_\nu^{num}(z)$ is the power along the polarization components $\nu=x,y$ in the simulations, and $N_\nu^{loc}(z)$ is given by Eq.(\ref{eq:N_nu_loc}). Note that the distance is a bounded quantity, $0 \le {\cal D}(z) \le 1$.

Figure~\ref{fig:loc_eq1} reports the spectra at two different propagation lengths, the evolution of the powers $N_{x,y}(z)$ during propagation, and the corresponding evolution of the distance to the RJ distributions ${\cal D}(z)$. Note the good agreement in (a)-(b) between the numerical spectra (blue line) and the local RJ distribution (orange line) over the truncated spectrum $|\omega| \le \omega_c^{loc}$. We stress in Fig.~\ref{fig:loc_eq1}(c) the remarkable agreement between $N_{x,y}^{loc}(z)$ computed from the local RJ equilibrium (\ref{eq:n_loc}) and the actual population $N_{x,y}^{num}(z)$ in the NLS simulation. It is important to note that the fitting procedure is based on minimizing the distance ${\cal D}(z)$, which means that nothing in the procedure explicitly enforces the agreement with the evolution of the individual polarization components, i.e., $N_\nu^{loc}(z)$ and $N_\nu^{num}(z)$ in NLS simulations, as shown in Fig.~\ref{fig:loc_eq1}(c). In addition, the small values of the distance to RJ equilibrium, remaining below ${\cal D} < 10^{-3}$ in panel (d), provide strong evidence for the validity of the local RJ equilibrium description of the spectra. 

\begin{figure}[t]
	\centering 
	\includegraphics[width=0.5\textwidth]{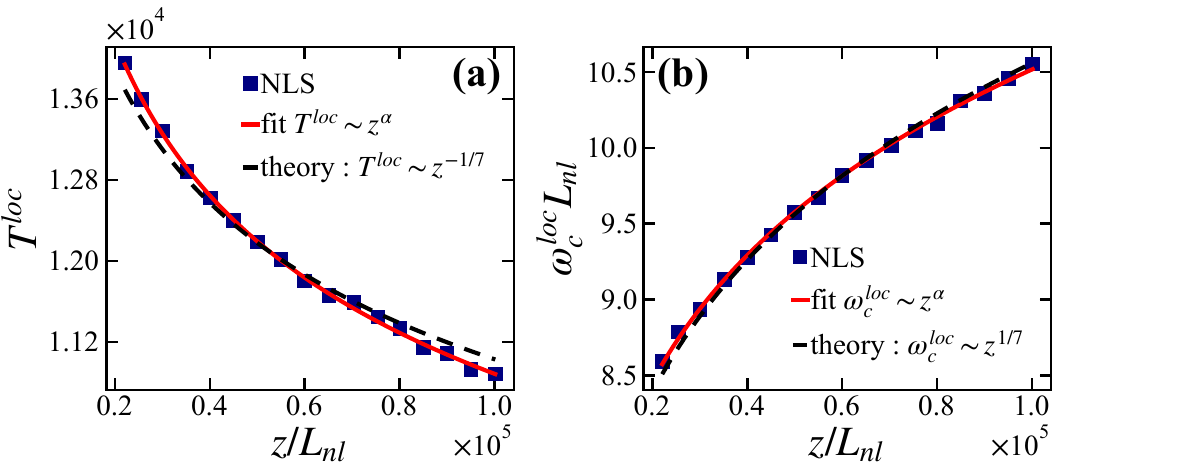}	
	\caption{\textbf{Scaling law for adiabatic cooling and spectral broadening in NLS simulations.} Evolution of the local temperature $T^{{loc}}$ (a), and local  frequency cut-off $\omega_c^{{loc}}$ (b), as a function of the propagation distance $z$. 
    Blue squares denote numerical data extracted from NLS simulations. The red curves are power-law fits of the form $z^{\alpha}$, yielding $\alpha \approx 0.135$ for $\omega_c^{{loc}}$, and $\alpha \approx -0.164$ for $T^{{loc}}$. These exponents are in good agreement with the theoretical predictions of Eq.(\ref{eq:scaling}), shown by the dashed black lines, $T^{{loc}}\sim z^{-1/7}$, and $\omega_c^{loc}\sim z^{1/7}$. The parameters $T^{loc}$ and $\omega_c^{loc}$ are obtained by fitting the NLS spectra with the local RJ equilibrium Eq.(\ref{eq:n_loc}),  following the procedure described in section~\ref{subsec:local_rj_eq}, see Fig.~\ref{fig:loc_eq1}.
    } 
	\label{fig:loc_eq2}
\end{figure}

\subsection{Theoretical scaling law for spectral broadening and adiabatic cooling}

The simulations show that, during the propagation in $z$, the bounded spectral window $[-{\omega}_{c}^{loc}(z),{\omega}_{c}^{loc}(z)]$ of the local RJ equilibrium expands, as shown in Fig.~\ref{fig:loc_eq1}(a)-(b). Consequently, the local temperature $T^{loc}$ and chemical potential $\mu^{loc}$ also display a $z$-dependence, as expected by imposing the conservation of $(N,E)$ in the presence of the $z$-dependent cut-offs truncation at $\omega_{c}^{loc}(z)$. 

We remark in particular that the local temperature $T^{loc}(z)$ plotted in Fig.~\ref{fig:loc_eq2}(a) exhibits a monotonic decrease. This is due to the continuous transfer (at a constant energy $E=$const) of the incoherent temporal fluctuations into the high-frequency tails of the spectrum distribution. This process can be described by the wave turbulence theory \cite{zanaglia26}. 
Indeed, there are terms in the kinetic equations that lead to an increase of the spectral bandwidth $\omega_c^{loc}(z)$.
Note in this respect that the incoherent collision term ${\cal C}oll_I[{\bm n}]$ in the kinetic Eqs.(\ref{eq:collx}-\ref{eq:colly}) does not contribute to a spectral broadening of the incoherent waves \cite{PRL10}.
Within the coherent collision term ${\cal C}oll_C[{\bm n}]$, only a single term contributes to the spectral broadening.  Considering the evolution of the $y$-polarization component, we obtain for $\omega > \omega_c^{loc}(z)$
\begin{eqnarray}
\partial_z n_y(\omega,z)=\frac{\rho^2 \gamma^2}{2\pi \beta} \int_{0}^\infty 
n_x \big(\omega-\sqrt{\zeta} \eta\big)
\, n_x\big(\omega-\sqrt{\zeta}/ \eta\big) 
\nonumber\\
\times \, n_y\big(\omega-\sqrt{\zeta} (\eta+1/\eta)\big) \, \frac{d\eta}{\eta},
\label{eq:kin_scal}
\end{eqnarray}
and we have a similar equation for $\omega < -\omega_c^{loc}(z)$. Then the width of the spectrum slowly increases, and we study this process by assuming $\omega_c^{loc}(z)$ large (i.e.,  $\omega_c^{loc}(z) \tau_0 \gg 1$). Using the local RJ equilibrium (\ref{eq:n_loc}), we get 
\begin{eqnarray}
\partial_z n_y(\omega,z)=\frac{\rho^2 \gamma^2}{2\pi \beta} \int_{0}^\infty 
 n_x^{loc} \big(\omega-\sqrt{\zeta} \eta\big)
\, n_x^{loc}\big(\omega-\sqrt{\zeta}/ \eta\big) 
\nonumber \\
\times \, n_y^{loc}\big(\omega-\sqrt{\zeta}( \eta+1/\eta) \big) 
\; {\bf 1}_{\eta_-<\eta<\eta_+} \,
\frac{d\eta}{\eta}
\label{eq:kin_to_powerlaw}
\end{eqnarray}
for $\omega \in [ \omega_c^{loc}(z), \omega_c^{loc}(z)+\sqrt{\zeta} ]$, with $\eta_-=(\omega-\omega_c^{loc})/\sqrt{\zeta}$ and $\eta_+=\sqrt{\zeta}/(\omega-\omega_c^{loc})$. The detailed evolution of  ${\omega}_{c}^{loc}(z)$ and the associated temperature $T^{loc}(z)$ depend on all physical parameters $\gamma, \rho, \beta$ and $\alpha$ that appear in the kinetic equation. Nevertheless, the evolution of the spectral front governed by Eq.(\ref{eq:kin_to_powerlaw}) and the associated local cut-off frequency $\omega_c^{loc}$ can be analyzed in detail \cite{zanaglia26}. 
We first note that energy conservation during propagation implies an inverse relationship between $T^{loc}(z)$ and $\omega_c^{loc}(z)$. Indeed, for large $\omega_c^{loc}$, the contribution of the chemical potential is negligible, and according to Eq.(\ref{eq:E_loc}) we have $E^{loc} =2 T^{loc} \omega_c^{loc}/\pi +\mu^{loc} N \simeq 2 T^{loc} \omega_c^{loc}/\pi \simeq$const, then leading to $T^{loc}(z) \sim 1/\omega_c^{loc}(z)$. 
Accordingly, the local RJ equilibrium scales as $n_\nu^{loc}(\omega) \sim T^{loc}/(\omega_c^{loc})^2 \sim 1/(\omega_c^{loc})^3$. Using this scaling in Eq.(\ref{eq:kin_to_powerlaw}) yields a theoretical power-law for the propagation of the spectral front in the asymptotic regime $z\gg L_{nl}$:
\begin{equation}
T^{loc}(z) \sim 1/\omega_c^{loc}(z) \sim 1/z^{1/7}.
\label{eq:scaling}
\end{equation}
The decrease in temperature accompanying spectral broadening thus results directly from energy conservation, which motivates the terminology ``adiabatic'' cooling.

The power-law scaling in Eq.(\ref{eq:scaling}) has been confirmed by the numerical simulations of the NLS equations. As discussed above through the Fig.~\ref{fig:loc_eq1}, the simulations show that the field evolves by following a RJ local equilibrium state (\ref{eq:n_loc}) during propagation. Figure~\ref{fig:loc_eq2} reports the corresponding local temperature $T^{loc}$, and local frequency cut-off $\omega_c^{loc}$, which are determined by following the procedure described in section~\ref{subsec:local_rj_eq}. The numerical results are fitted by power laws $\sim z^\alpha$, with exponents $\alpha=0.135$ for $\omega_c^{loc}(z)$, and $\alpha=0.164$ for $T^{loc}(z)$, in reasonable agreement with the predicted value $1/7\sim 0.143$ in Eq.(\ref{eq:scaling}).

\begin{figure}
	\centering 
	\includegraphics[width=0.5\textwidth]{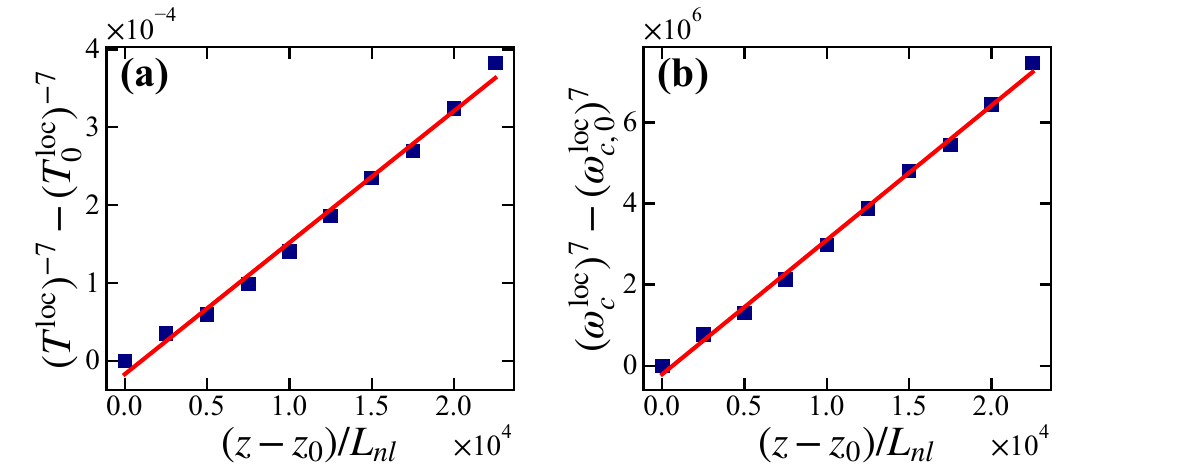}	
	\caption{{\bf Scaling law for adiabatic cooling and spectral broadening in simulations of the kinetic equation.}
    Numerical integration of the kinetic Eqs.(\ref{eq:kin_n_x}–\ref{eq:kin_n_y}) starting from a local equilibrium initial state. The evolution during propagation of the local temperature and local frequency cut-off (squares) follows the power-law behavior predicted by the theory in Eq.(\ref{eq:scaling}), $(T_c^{loc})^{-7} \sim (\omega_c^{loc})^{7} \sim z$. The solid red lines show the linear scaling behavior $\propto z$. Parameters: $\alpha L_{nl} = 1$ and $\omega_{c,0}^{loc} \tau_0 = 10$; the initial condition is close to a local RJ spectrum characterized by the reference values $T_0^{loc}$ and $\omega_{c,0}^{loc}$, with rapidly decaying tails for $|\omega| > \omega_{c,0}^{loc}$ (see the text for details).
    }
	\label{fig:scalin_law_cin}
\end{figure}

We finally note that the scaling law in Eq.(\ref{eq:scaling}) is also confirmed by direct numerical simulations of the kinetic Eqs.(\ref{eq:kin_n_x}–\ref{eq:kin_n_y}). However, the numerical integration of the kinetic equations becomes increasingly demanding at long propagation distances, so that we were unable to obtain sufficiently accurate solutions over the range of propagation lengths explored for the NLS equation in Fig.~\ref{fig:loc_eq2}. Accordingly, rather than simulating the initial transient leading to the local equilibrium, we directly initialize the system in a local equilibrium state with rapidly decaying spectral tails for $|\omega| > \omega_{c,0}^{loc}$. The corresponding initial values of the temperature $T_0^{loc}$, and frequency cut-off $\omega_{c,0}^{loc}$, therefore serve only as reference values. Figure~\ref{fig:scalin_law_cin} shows the subsequent evolution of the temperature and frequency cut-off, obtained from the simulation of the kinetic Eqs.(\ref{eq:kin_n_x}–\ref{eq:kin_n_y}). The numerical results show that the system follows a local RJ equilibrium state given by Eq.(\ref{eq:n_loc}), whose local temperature and local frequency cut-off were determined by following the same procedure described above for the NLS simulations, see section~\ref{subsec:local_rj_eq}. The results reported in Fig.~\ref{fig:scalin_law_cin} confirm the power-law scaling predicted theoretically in Eq.(\ref{eq:scaling}).

\section{Persistent increase of the degree of polarization}

We have seen in the previous section that the propagation of the incoherent optical wave is characterized by a process of adiabatic cooling in which temporal fluctuations are continuously transferred toward the high-frequency tails of a local RJ equilibrium state. In the following, we show that this spectral broadening of the waves is responsible for a persistent increase in the degree of polarization.

We consider an initial condition characterized by two uncorrelated random waves with Gaussian-shaped initial spectra, see Eq.(\ref{eq:init_gauss}) with $N_x^0=N_y^0$. The initial condition fixes the total power and energy $(N,E)$, which are conserved quantities throughout the propagation. As discussed in the previous section, the field converges toward a local RJ equilibrium state (\ref{eq:n_loc}), which is characterized by the three local parameters $(T^{loc},\mu^{loc},\omega_c^{loc})$. According to the power-law  scaling (\ref{eq:scaling}), the frequency cut-off increases during propagation. For each $\omega_c$, we use Eqs.(\ref{eq:EsN_loc}-\ref{eq:E_loc}) to compute the local temperature and chemical potential $(T^{loc}, \mu^{loc})$ from the conserved quantities $(N,E)$ -- the chemical potential $\mu^{loc}$ is determined by (\ref{eq:EsN_loc}) with $E^{loc}=E$ and $N^{loc}=N$, and then $T^{loc}$ is fixed by (\ref{eq:E_loc}).

Once $(T^{loc},\mu^{loc})$ are determined for a given value of $\omega_c^{loc}$, we calculate the corresponding local powers $N_\nu^{loc}$ for the two polarization components ($\nu=x,y$) by using Eq.(\ref{eq:N_nu_loc}). Finally, the degree of polarization is determined by these local powers in the absence of correlations, ${\cal P}=(1 - 4 N_x^{loc} N_y^{loc}/N^2 )^{1/2}$. 

\begin{figure}
	\centering 
	\includegraphics[width=0.5\textwidth]{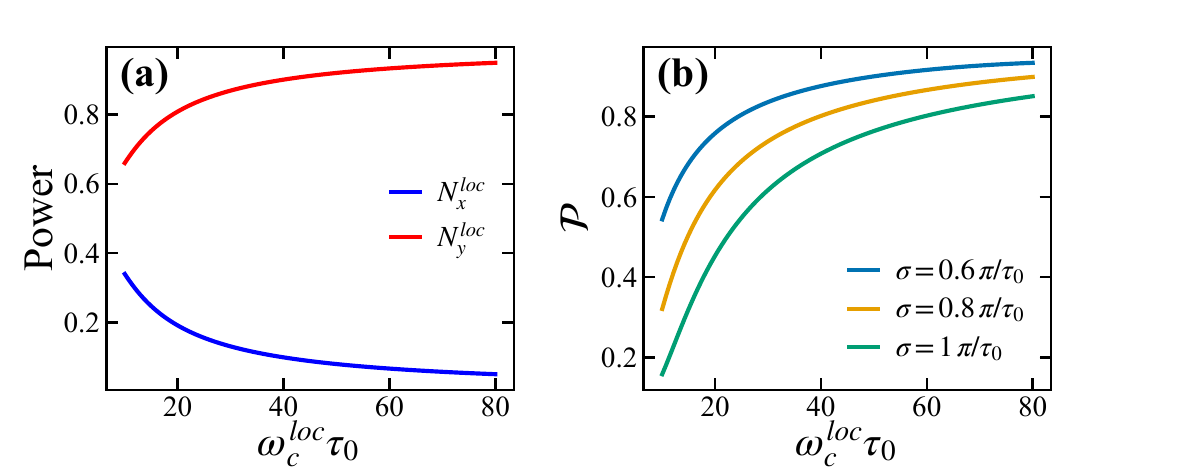}	
	\caption{\textbf{Persistent growth of the degree of polarization toward unity.} Optical powers $N^{loc}_x$ (blue curve) and $N^{loc}_y$ (red curve) (from Eq.(\ref{eq:N_nu_loc})) (a), and corresponding degree of polarization ${\cal P}$ (b), as a function of the local frequency cut-off $\omega^{loc}_{c}$. In panel (b), different curves correspond to different initial spectral widths $\sigma$. As $\omega^{loc}_c$ increases (corresponding to longer propagation distances $z$), the $y-$polarization axis becomes more populated and ${\cal P}$ increases toward unity, indicating that self-polarization progresses continuously along with spectral broadening. Narrower initial spectral widths (i.e., stronger nonlinear effects) tend to accelerate the increase of ${\cal P}$ toward unity. Parameters: $\alpha  L_{nl} = 1$ and $\kappa=2/3$. In (a),  $\sigma = 0.8\pi/\tau_0$; in (b), the corresponding values of $\sigma$ are indicated in the legend. 
    } 
	\label{fig:virt}
\end{figure}

Figure~\ref{fig:virt} presents the evolution of the powers $N_\nu^{loc}$ with $\omega_c^{loc}$ in panel (a), and the corresponding evolution of the degree of polarization ${\cal P}$ in panel (b), for various initial spectral widths $\sigma$.
We note that the $y$-polarization axis gradually becomes more populated at the expense of the $x$-axis as $\omega_c^{loc}$ increases. Panel (b) shows that the degree of polarization ${\cal P}$ increases toward unity, regardless of the initial spectral widths $\sigma$. Note however that the rate of self-polarization depends on the initial spectral width. These results show that self-polarization progresses continuously with increasing $\omega_c^{loc}$, corresponding to the progressive broadening of the local equilibrium spectrum during propagation according to the scaling (\ref{eq:scaling}).

\section{Discussion and conclusion}

By considering the coherently coupled NLS equations governing light propagation in weakly birefringent optical fibers, we have characterized a nonequilibrium process of  self-polarization driven by the natural thermalization of the optical field, recently observed experimentally \cite{torres26}. Within the framework of wave turbulence theory, we have provided a detailed description of this nonequilibrium process. The analysis of the kinetic equations unveils the mechanism that triggers the initial stage of light self-polarization. In a second stage, we show that self-polarization proceeds through an adiabatic cooling mechanism, whereby polarization fluctuations are progressively transferred toward the high-frequency tails of the spectrum. The optical field therefore follows a local RJ equilibrium state characterized by a decreasing local temperature, resulting in an increasing degree of polarization during propagation. Since no high-frequency cutoff is available to regularize the ultraviolet catastrophe, the self-polarization process appears to be intrinsically unbounded, with the degree of polarization continuously increasing toward unity. In other words, the conservative classical dynamics does not provide an intrinsic upper limit to the transfer of spectral energy toward increasingly high frequencies, so that the associated adiabatic cooling can, in principle, proceed without saturation. 

It is important to emphasize, however, that this conclusion must be understood within the domain of validity of the classical and conservative NLS model considered in this work. In particular, the model neglects material absorption and higher-order dissipative nonlinearities, which are expected to become increasingly relevant as the spectral content extends toward higher frequencies. Such effects would provide additional mechanisms for limiting the effect of self-polarization in an actual experiment. 

Moreover, the classical description is ultimately expected to break down at sufficiently high frequencies, where the photon-number occupation drops below unity and the discreteness of the field quanta starts to matter. In this regime, the spectral distribution should approach a Bose–Einstein form \cite{zanaglia26b}, thereby introducing an ultimate physical constraint on the adiabatic cooling process. The emergence of this quantum-statistical regime provides a natural limitation to the otherwise unbounded behavior predicted by the classical conservative theory. It is worth emphasizing that the characteristic power-law scaling underlying adiabatic cooling is intrinsically tied to the classical description, further highlighting the classical nature of this phenomenon. We anticipate that the ultimate quantum limitation will become relevant only at extremely large propagation distances. Indeed, because the nonlinearities that characterize typical nonlinear media are relatively weak, the spectral evolution toward the very high-frequency regime is expected to be exceedingly slow, requiring propagation lengths far beyond those generally accessible in practical settings \cite{chiocchetta16}. Thus, over experimentally relevant distances and within the regime where the classical conservative description remains applicable, the self-polarization process can be regarded as effectively unbounded.

We finally note that work is currently in progress to provide an experimental signature of the formation of a local RJ equilibrium state underlying the self-polarization process. The principal difficulty lies in the long propagation distances required for the adiabatic cooling stage to develop, as predicted by the weak-turbulence analysis presented in this work. It should be emphasized, however, that these large length scales are a direct consequence of the weakly nonlinear regime considered here, which was specifically chosen to enable a theoretical description within the framework of wave turbulence theory. As discussed in section~\ref{subsec:repolar_strong}, the dynamics becomes dramatically faster beyond this regime. While the theoretical description of adiabatic cooling beyond the weakly nonlinear regime remains to be developed, the substantial acceleration of the evolution suggests that the corresponding length scales could be significantly reduced. In this context, the experimental methodology recently implemented in Ref.\cite{torres26} may provide a useful framework for investigating the possible emergence of nonequilibrium self-polarization and its associated cooling dynamics.

\section*{Acknowledgements}
Funding was provided by Agence Nationale de la Recherche (Grants No. ANR-23-CE30-0021, No. ANR-21-ESRE-0040). Calculations were performed using HPC resources from DNUM-CCUB (Universit\'e Bourgogne Europe). I.C. acknowledges financial support from the PE0000023-NQSTI project by the Italian Ministry of University and Research, co-funded by the European Union - NextGeneration EU; from Provincia Autonoma di Trento (PAT), partly via the Q@TN initiative. L.Z. and M.F. acknowledge funding from the Italian Ministry of University and Research (MUR) through the Italian Science Fund (FIS2), project SOFT (H53C24001530001). S.W. acknowledges support from the European Research Council, project MULTIPROBE (101224071).


\appendix

\begin{strip}
\section{Derivation of the wave turbulence kinetic Eqs.(\ref{eq:kin_n_x}-\ref{eq:kin_n_y})}
\label{app:kinetic}

The kinetic equations can be derived from the NLS Eqs.(\ref{eq:nlsu_x}-\ref{eq:nlsu_y}) by using the standard wave turbulence theory \cite{zakharov92,Newell01,zakharov04,nazarenko11}. The kinetic equations can be written in the following general form
\begin{eqnarray}
\partial_z n_{x}(\omega,z) &= \kappa^2 {\cal C}oll_I[n_{x},n_{y}] + \rho^2 {\cal C}oll_C[n_{x},n_{y}], 
\label{eq:collx_app}
\\
\partial_z n_{y}(\omega,z) &= \kappa^2 {\cal C}oll_I[n_{y},n_{x}] + \rho^2 {\cal C}oll_C[n_{y},n_{x}],
\label{eq:colly_app}
\end{eqnarray}
where the optical spectra $n_\nu(\omega,z)$ along the orthogonal polarization components $\nu=x,y$ are defined from $\left< {\tilde u}_\nu(\omega_1,z) {\tilde u}_\nu^*(\omega_2,z) \right>=2\pi n_\nu(\omega_1,z) \delta(\omega_1-\omega_2)$ under the assumption of stationary statistics, with  ${\tilde u}_\nu(\omega,z)=\int u_\nu(t,z) \exp(-i\omega t) dt$ the Fourier transform of the wave amplitude $u_\nu(t,z)$. In Eqs.(\ref{eq:collx_app}-\ref{eq:colly_app}), ${\cal C}oll_I[{\bm n}]$ and ${\cal C}oll_C[{\bm n}]$ denote, respectively, the collision terms originating from the incoherent and coherent terms in the NLS Eqs.(\ref{eq:nlsu_x}-\ref{eq:nlsu_y}). The incoherent collision term reads:
\begin{eqnarray}
&&\hspace*{-0.2in}
{\cal C}oll_I[n_{x},n_{y}] = \frac{ \gamma^2}{2\pi} \iiint d\omega_{1-3} \,  
 n_{x}(\omega_1) \, n_{y}(\omega_2)\,n_{y}(\omega_3)\,n_{x}(\omega)  \left( n_{x}^{-1}(\omega)+n_{y}^{-1}(\omega_3)-n_{y}^{-1}(\omega_2)-n_{x}^{-1}(\omega_1)\right) \nonumber\\ 
&&\hspace*{2.3in}
\times \, \delta\big(k_x(\omega_1)+k_y(\omega_2)-k_y(\omega_3)-k_x(\omega)\big) \
\delta(\omega_1+\omega_2-\omega_3-\omega), 
\label{eq:wtcollu}\\
&&\hspace*{-0.2in}
{\cal C}oll_I[n_{y},n_{x}] = \frac{ \gamma^2}{2\pi} \iiint d\omega_{1-3} \,  
 n_{y}(\omega_1) \, n_{x}(\omega_2)\,n_{x}(\omega_3)\,n_{y}(\omega)  \left( n_{y}^{-1}(\omega)+n_{x}^{-1}(\omega_3)-n_{x}^{-1}(\omega_2)-n_{y}^{-1}(\omega_1)\right) \nonumber \\ 
&&\hspace*{2.3in}
\times \, \delta\big(k_y(\omega_1)+k_x(\omega_2)-k_x(\omega_3)-k_y(\omega)\big) \
\delta(\omega_1+\omega_2-\omega_3-\omega),
\label{eq:wtcollv}
\end{eqnarray}
with $d\omega_{1-3}=d\omega_1 d\omega_2 d\omega_3$.
The collision term for the coherent interaction reads
\begin{eqnarray}
&&\hspace*{-0.2in}
{\cal C}oll_C[n_{x},n_{y}] = \frac{ 2\gamma^2}{2\pi} \iiint d\omega_{1-3} \,  
 n_{y}(\omega_1) \, n_{y}(\omega_2)\,n_{x}(\omega_3)\,n_{x}(\omega)  \left( n_{x}^{-1}(\omega)+n_{x}^{-1}(\omega_3)-n_{y}^{-1}(\omega_2)-n_{y}^{-1}(\omega_1)\right) \nonumber\\ 
&&\hspace*{2.3in}
\times \, \delta\big(k_y(\omega_1)+k_y(\omega_2)-k_x(\omega_3)-k_x(\omega)\big) \
\delta(\omega_1+\omega_2-\omega_3-\omega), 
\label{eq:wtcollu}\\
&&\hspace*{-0.2in}
{\cal C}oll_C[n_{y},n_{x}] = \frac{ 2\gamma^2}{2\pi} \iiint d\omega_{1-3} \,  
 n_{x}(\omega_1) \, n_{x}(\omega_2)\,n_{y}(\omega_3)\,n_{y}(\omega)  \left( n_{y}^{-1}(\omega)+n_{y}^{-1}(\omega_3)-n_{x}^{-1}(\omega_2)-n_{x}^{-1}(\omega_1)\right) \nonumber \\ 
&&\hspace*{2.3in}
\times \, \delta\big(k_x(\omega_1)+k_x(\omega_2)-k_y(\omega_3)-k_y(\omega)\big) \
\delta(\omega_1+\omega_2-\omega_3-\omega).
\label{eq:wtcollv}
\end{eqnarray}
Note that, because of the degenerate resonances inherent to the purely one-dimensional problem considered here, the collision term describing the self-interaction (first nonlinear term in Eqs.(\ref{eq:nlsu_x}-\ref{eq:nlsu_y})) vanishes identically, see Refs.\cite{zakharov92,Newell01,zakharov04,nazarenko11}.
Two integrals can be computed owing to the Dirac $\delta-$functions, which gives:
\begin{eqnarray}
&&\hspace*{-0.2in}
\partial_z n_{x}(\omega,z) = \frac{\kappa^2 \gamma^2}{4\pi \beta} \int  
 n_{x}({\omega'}) \, n_{y}(\omega)\,n_{y}({\omega'})\,n_{x}(\omega)  \left( n_{x}^{-1}(\omega)+n_{y}^{-1}({\omega'})-n_{y}^{-1}(\omega)-n_{x}^{-1}({\omega'})\right)  \frac{d{\omega'}}{|{\omega'}-\omega|} \nonumber\\
&&\hspace*{0.3in}
+ \frac{2 \rho^2 \gamma^2}{4\pi \beta} \int n_{y}(\omega +
 {\omega'}) \, n_{y}(\omega-\zeta/{\omega'})\,n_{x}(\omega +{\omega'}-\zeta/{\omega'})\,n_{x}(\omega) \nonumber\\ 
&&\hspace*{2.2in}
 \times \left( n_{x}^{-1}(\omega)+n_{x}^{-1}(\omega  + {\omega'}-\zeta/{\omega'})-n_{y}^{-1}(\omega +{\omega'})-n_{y}^{-1}(\omega-\zeta/{\omega'})\right) \frac{d{\omega'}}{|{\omega'}|} , \\ 
&&\hspace*{-0.2in}
\partial_z n_{y}(\omega,z) = \frac{\kappa^2 \gamma^2}{4\pi \beta} \int  
 n_{x}({\omega'}) \, n_{y}(\omega)\,n_{y}({\omega'})\,n_{x}(\omega)  \left( n_{x}^{-1}(\omega)+n_{y}^{-1}({\omega'})-n_{y}^{-1}(\omega)-n_{x}^{-1}({\omega'})\right)  \frac{d{\omega'}}{|{\omega'}-\omega|} \nonumber\\
&&\hspace*{0.3in} +\frac{2 \rho^2 \gamma^2}{4\pi \beta} \int   
 n_{x}(\omega +{\omega'}) \, n_{x}(\omega + \zeta/{\omega'})\,n_{y}(\omega +{\omega'} +
 \zeta/{\omega'})\,n_{y}(\omega) \nonumber\\ 
&&\hspace*{2.2in}
 \times  \left( n_{y}^{-1}(\omega)+n_{y}^{-1}(\omega +{\omega'} +\zeta/{\omega'})-n_{x}^{-1}(\omega +{\omega'})-n_{x}^{-1}(\omega +\zeta/{\omega'})\right) \frac{d{\omega'}}{|{\omega'}|} ,
\end{eqnarray}
where $\zeta = 2 \alpha/\beta >0$. With the change of variable $\eta=\omega'/\sqrt{\zeta}$ in the second coherent collision term, the kinetic equations take the form:
\begin{eqnarray}
&&\hspace*{-0.2in}
\partial_z n_{x}(\omega,z) = \frac{\kappa^2 \gamma^2}{4\pi \beta} \int  
 n_{x}({\omega'}) \, n_{y}(\omega)\,n_{y}({\omega'})\,n_{x}(\omega)  \left( n_{x}^{-1}(\omega)+n_{y}^{-1}({\omega'})-n_{y}^{-1}(\omega)-n_{x}^{-1}({\omega'})\right)  \frac{d{\omega'}}{|{\omega'}-\omega|} \nonumber\\
&&\hspace*{0.3in}+ \frac{2 \rho^2 \gamma^2}{4\pi \beta} \int   
 n_{y}(\omega + \sqrt{\zeta} {\eta}) 
 \, n_{y}(\omega-\sqrt{\zeta}/{\eta} )\,n_{x}(\omega + \sqrt{\zeta} {\eta}-\sqrt{\zeta}  /{\eta} )\,n_{x}(\omega) 
 \nonumber \\ 
&&\hspace*{2in}
\times \left( n_{x}^{-1}(\omega)+n_{x}^{-1}(\omega +\sqrt{\zeta} {\eta}- \sqrt{\zeta}  / {\eta})-n_{y}^{-1} (\omega +\sqrt{\zeta} {\eta})-n_{y}^{-1}(\omega-\sqrt{\zeta}  / {\eta})\right) \frac{d{\eta}}{|{\eta}|} , \label{eq:kin_n_x_app}\\ 
&&\hspace*{-0.2in}
\partial_z n_{y}(\omega,z) = \frac{\kappa^2 \gamma^2}{4\pi \beta} \int  
 n_{x}({\omega'}) \, n_{y}(\omega)\,n_{y}({\omega'})\,n_{x}(\omega)  \left( n_{x}^{-1}(\omega)+n_{y}^{-1}({\omega'})-n_{y}^{-1}(\omega)-n_{x}^{-1}({\omega'})\right)  \frac{d{\omega'}}{|{\omega'}-\omega|} \nonumber\\
&&\hspace*{0.3in}+\frac{2 \rho^2 \gamma^2}{4\pi \beta} \int   
 n_{x}(\omega + \sqrt{\zeta} {\eta}) \, n_{x}(\omega + \sqrt{\zeta} / {\eta})\,n_{y}(\omega + \sqrt{\zeta} {\eta} +
\sqrt{\zeta}  /{\eta} )\,n_{y}(\omega) \nonumber\\ 
&&\hspace*{2in}
\times  \left( n_{y}^{-1}(\omega)+n_{y}^{-1}(\omega +\sqrt{\zeta} {\eta} +\sqrt{\zeta} / {\eta})
 -n_{x}^{-1}(\omega + \sqrt{\zeta} {\eta})-n_{x}^{-1}(\omega +\sqrt{\zeta}  /{\eta} )\right) \frac{d{\eta}}{|{\eta}|} .
\label{eq:kin_n_y_app}
\end{eqnarray}

\section{Derivation of the kinetic equations for the optical powers, Eqs.(\ref{eq:dzNx}-\ref{eq:dzNy}) and Eq.(\ref{eq:init_stage})}
\label{app:kinetic_power}

To derive the kinetic equations for the optical powers (\ref{eq:dzNx}-\ref{eq:dzNy}), we start by remarking that the singularity in the second coherent collision terms in (\ref{eq:kin_n_x_app}-\ref{eq:kin_n_y_app}) can be removed by the change of variable $\eta \to s e^\theta$, $s=\pm 1$, so that the kinetic equations take the form 
\begin{eqnarray}
&&\hspace*{-0.2in}
\partial_z n_{x}(\omega,z) = 
\frac{\kappa^2 \gamma^2}{4\pi \beta} \int  
 n_{x}({\omega'}) \, n_{y}(\omega)\,n_{y}({\omega'})\,n_{x}(\omega)  \left( n_{x}^{-1}(\omega)+n_{y}^{-1}({\omega'})-n_{y}^{-1}(\omega)-n_{x}^{-1}({\omega'})\right)  \frac{d{\omega'}}{|{\omega'}-\omega|} \nonumber\\
&&\hspace*{0.3in}
+ \frac{2\rho^2 \gamma^2}{4\pi \beta} \sum_s 
\int   
 n_{y}(\omega + \sqrt{\zeta} s e^\theta) 
 \, n_{y}(\omega-\sqrt{\zeta} s e^{-\theta} )\,n_{x}(\omega + \sqrt{\zeta} s e^\theta-\sqrt{\zeta}  s e^{-\theta} )\,n_{x}(\omega) \nonumber\\ 
&&\hspace*{1.5in}
\times \left( n_{x}^{-1}(\omega)+n_{x}^{-1}(\omega +\sqrt{\zeta} s e^\theta - \sqrt{\zeta}  s  e^{-\theta})-n_{y}^{-1}(\omega +\sqrt{\zeta} s  e^\theta )-n_{y}^{-1}(\omega-\sqrt{\zeta}  s  e^{-\theta})\right) d\theta  ,
\label{eq:red_wtkin_x} 
\\ 
&&\hspace*{-0.2in}
\partial_z n_{y}(\omega,z) = 
\frac{\kappa^2 \gamma^2}{4\pi \beta} \int   
 n_{y}({\omega'}) \, n_{x}(\omega)\,n_{x}({\omega'})\,n_{y}(\omega)  \left( n_{y}^{-1}(\omega)+n_{x}^{-1}({\omega'})-n_{x}^{-1}(\omega)-n_{y}^{-1}({\omega'})\right) \frac{d{\omega'}}{|{\omega'}-\omega|} \nonumber\\ 
&&\hspace*{0.3in}
+\frac{2 \rho^2 \gamma^2}{4\pi \beta} \sum_s  \int   
 n_{x}(\omega + \sqrt{\zeta} s e^{\theta}  ) \, n_{x}(\omega + \sqrt{\zeta}  s e^{-\theta} )\,n_{y}(\omega + \sqrt{\zeta} s e^{\theta}  +
\sqrt{\zeta}  s e^{-\theta}  )\,n_{y}(\omega) 
\nonumber\\ 
&&\hspace*{1.5in}
\times \left( n_{y}^{-1}(\omega)+n_{y}^{-1}(\omega +\sqrt{\zeta} s  e^{\theta}+\sqrt{\zeta}  s e^{-\theta} )
 -n_{x}^{-1}(\omega + \sqrt{\zeta} s e^{\theta} )-n_{x}^{-1}(\omega +\sqrt{\zeta}  s e^{-\theta}  )\right) d\theta.
\label{eq:red_wtkin_y} 
\end{eqnarray}
The wave turbulence kinetic equations for the optical power $N_\nu = \frac{1}{2\pi} \int n_\nu(\omega,z) d\omega$ are obtained by integrating over $\omega$ the kinetic Eqs.(\ref{eq:red_wtkin_x}-\ref{eq:red_wtkin_y}) governing the evolution of the spectra $n_\nu(\omega,z)$. As commented in the text, we first note that the incoherent collision term ${\cal C}oll_I[{\bm n}]$ does not contribute to a power exchange among the polarization components, as revealed by a permutation of the variables $\omega$ and $\omega'$ leading to $\int  {\cal C}oll_I[{\bm n}]d\omega=0$. Integrating the coherent collision term $\int {\cal C}oll_C[{\bm n}] d\omega$ and performing the change of variable  $\omega \to \omega - \sqrt{\zeta} \sinh\theta$ leads to equations for the optical powers
\begin{eqnarray}
&&\hspace*{-0.2in}
\partial_z N_{x}(z) = \frac{2\rho^2 \gamma^2}{(2\pi)^2 \beta}  \int d\omega  \int d\theta 
 n_{y}(\omega+\sqrt{\zeta}  \cosh \theta) \, n_{y}(\omega -\sqrt{\zeta}  \cosh \theta)\,n_{x}(\omega +\sqrt{\zeta}  \sinh \theta)\,n_{x}(\omega-\sqrt{\zeta}  \sinh \theta) \nonumber\\ 
&&\hspace*{1.2in}
\times \left[ n_{x}^{-1} \big(\omega +\sqrt{\zeta}  \sinh \theta \big)+n_{x}^{-1} \big(\omega -\sqrt{\zeta}  \sinh \theta\big)-n_{y}^{-1}\big(\omega +\sqrt{\zeta}  \cosh \theta\big)-n_{y}^{-1} \big(\omega - \sqrt{\zeta}  \cosh \theta \big)\right] ,
\label{eq:dzNx_app} \\ 
&&\hspace*{-0.2in}
\partial_z N_{y}(z) =  - \partial_z N_{x}(z)   .
\label{eq:dzNy_app}
\end{eqnarray}
To derive Eq.(\ref{eq:init_stage}) governing the initial stage of the repolarization process, we consider Gaussian-shaped initial spectra $n_{\nu}^0(\omega)= \sqrt{2\pi} N_{\nu}^0 \exp\big(-\omega^2/(2\sigma^2)\big)/\sigma$, for $\nu=x,y$.
Substituting these expressions into Eq.(\ref{eq:dzNx_app}), and computing the integral over $\omega$ leads to
\begin{eqnarray}
\partial_z N_{x}(z) = \frac{4\rho^2 \gamma^2}{\sqrt{3} \beta \sigma^2} N_x^0 N_y^0 
\int  d\theta \Big[ N_y^0 e^{- \frac{\zeta}{\sigma^2}(\frac{1}{3} \sinh^2\theta+\cosh^2\theta)}  
- N_x^0 e^{- \frac{\zeta}{\sigma^2}(\sinh^2\theta+\frac{1}{3} \cosh^2\theta)}\Big].
\end{eqnarray}
The integration over $\theta$ can be expressed in terms of the modified Bessel function of the
second kind $K_0(z)=\int_0^\infty e^{-z \cosh(t)} dt$, for $z>0$. Using the identities $\cosh^2\theta=\frac{1}{2}(\cosh2\theta + 1)$ and $\sinh^2\theta=\frac{1}{2}(\cosh2\theta - 1)$, we have $\int_{-\infty}^{\infty} e^{-z \cosh^2\theta} d\theta=e^{-z/2} K_0(z/2)$, $\int_{-\infty}^{\infty} e^{-z \sinh^2\theta} d\theta=e^{z/2} K_0(z/2)$. In this way we obtain
\begin{eqnarray}
\int_{-\infty}^{\infty} e^{-\frac{\zeta}{\sigma^2} (\frac{1}{3} \sinh^2\theta+\cosh^2\theta)} d\theta=e^{-\frac{\zeta}{3\sigma^2}} K_0\Big(\frac{2 \zeta}{3\sigma^2}\Big), \qquad \int_{-\infty}^{\infty} e^{-\frac{\zeta}{\sigma^2} (\sinh^2\theta+\frac{1}{3} \cosh^2\theta)} d\theta=e^{\frac{\zeta}{3\sigma^2}} K_0 \Big(\frac{2 \zeta}{3\sigma^2}\Big),    
\end{eqnarray}
which finally gives Eq.(\ref{eq:init_stage})
\begin{eqnarray}
\partial_z N_{x}(z) = \frac{4\rho^2 \gamma^2}{\sqrt{3} \beta \sigma^2} N_y^0 N_x^0 e^{-\zeta/(3\sigma^2)}  K_0 \Big( \frac{2\zeta}{3\sigma^2}\Big) \, \Big( N_y^0-N_x^0 e^{2\zeta/(3\sigma^2)}\Big) .
\label{eq:init_stage_app}
\end{eqnarray}    

\end{strip}



\begin{thebibliography}{00}
\bibitem{note}
It is worth noting that this condition also ensures that the nonlinear dispersion branches remain well separated from each other. Indeed, the nonlinear corrections modify the dispersion relations as  $k_x^{nl}(\omega)=\beta \omega^2 + \alpha +\gamma (2 N_x + \kappa N_y)$ and $k_y^{nl}(\omega)=\beta \omega^2 - \alpha +\gamma (2 N_y + \kappa N_x)$. The requirement $k_x^{nl}(\omega) > k_y^{nl}(\omega)$ yields $\alpha L_{nl} > (1-\kappa/2) (N_y-N_x)/N$, where $L_{nl}=1/(\gamma N)$ is the nonlinear length.
\end{thebibliography}


\end{document}